\documentclass[12pt, reqno]{article}
\usepackage{amssymb,amsmath,latexsym}
\usepackage{fancyhdr}
\usepackage{stmaryrd}
\usepackage{rotating}
\usepackage{setspace}
\usepackage{hyperref}
\usepackage{graphicx}
\usepackage{booktabs}
\usepackage{dcolumn}          
\usepackage{bm}               
\usepackage[affil-it]{authblk}

\newcommand {\rb} {\right]}
\newcommand {\lb} {\left[}
\newcommand {\lp} {\left(}
\newcommand {\rp} {\right)}
\newcommand {\lt}  {\Lambda}
\newcommand {\Li}  {\text{Li}}
\newcommand {\G}  {\Gamma}
\newcommand {\g}  {\gamma}
\newcommand {\T}  {\tau}
\newcommand {\fpi} {{16 \pi^2}}
\newcommand {\n}  {{\eta(q)}}    
\newcommand {\w}  {\omega}
\newcommand {\mc}[1]{\mathcal{#1}}
\newcommand {\mf}[1]{\mathfrak{#1}}  
\newcommand {\I} {\mathcal{I}}     
\newcommand {\B} {\mathcal{B}}     
\newcommand {\CC} {{\mathbb C}}   
\newcommand {\RR} {{\mathbb R}}   
\newcommand {\ZZ} {{\mathbb Z}}   
\newcommand {\ii} {{\text{i}}}   
\newcommand {\diff} {{\rm d}} 
\newcommand {\tr} {{\rm tr}}  
\newcommand {\Tr} {{\rm Tr}}  
\newcommand {\rep}[1]{\mathbf{(#1)}}
\newcommand {\rank}{\text{rank}}
\newcommand {\del}{\partial}
\newcommand {\delbar}{\bar{\partial}}
\newcommand {\z}{\vec{z}}
\newcommand {\tht} {\vartheta}

\date{}

\begin{document}

\title{\hfill \vbox{\normalsize \hbox{EFI-13-25}} \bigskip \bigskip \bigskip \\ 
Jacobi Forms of Higher Index and Paramodular Groups in $\mathcal{N}=2$, $D=4$ Compactifications of String Theory}

\author{
    Caner Nazaroglu%
    \thanks{\texttt{cnazaroglu@uchicago.edu}}
    }
\affil{Enrico Fermi Institute and Department of Physics, University of Chicago \\
	5640 S. Ellis Avenue, Chicago, IL 60637, USA}
\maketitle
\bigskip
\bigskip

\begin{center}
\bf{Abstract}
\end{center}
\bigskip
We associate a Jacobi form
over a rank $s$ lattice to $\mathcal{N}=2$, $D=4$ heterotic string compactifications 
which have $s$ Wilson lines at a generic point in the 
vector multiplet moduli space. Jacobi forms of index $m=1$ and $m=2$  have appeared earlier in the
context of threshold corrections to heterotic string couplings. We emphasize that higher index Jacobi forms as well as Jacobi forms
of several variables over more
generic even lattices also appear and construct models in which they arise.
In particular, we construct an orbifold model which can be connected to models that give index $m=3$, $4$ or $5$ Jacobi forms
through the Higgsing process.
Constraints from being a Jacobi form are then employed 
to get threshold corrections using only  partial information on the spectrum. We apply this procedure for  index
$m=3$, $4$ or $5$ Jacobi form examples and also for Jacobi forms over $A_2$ and $A_3$ root lattices.
Examples with a single Wilson line are examined in detail and we display the relation of Siegel forms over a paramodular 
group $\Gamma_m$ to these models, where $\Gamma_m$ is associated with the T-duality group of the models we study.
Finally, results on the heterotic string side
are used to clarify the linear mapping of vector multiplet moduli to Type IIA duals without using the one-loop cubic part of 
the prepotential on the Type II side,
and also to give predictions for the geometry of the dual
Calabi-Yau manifolds.

\bigskip

\pagebreak

\tableofcontents

\section{Introduction}
A variety of modular forms and Jacobi forms arise  frequently in string theory. 
Constraints that come from having
a modular or Jacobi form in a physical quantity are highly nontrivial and may even 
allow us to get information not accessible through direct methods.
The focus in this work will be on $D=4$, $\mc{N}=2$ string compactifications and their relation to Jacobi forms of possibly 
several variables.

On the heterotic string side, the modified index of \cite{Cecotti:1992qh} is used to compute threshold corrections to gauge
couplings and gravitational couplings \cite{Antoniadis:1992rq}. 
In the case where no Wilson line is turned on, this index gives rise to a modular form \cite{Harvey:1995fq}. The case with 
Wilson lines, on the other hand, involves Jacobi forms. In \cite{LopesCardoso:1996nc},
index $m=1$ Jacobi forms
in the sense of \cite{Zagier} are found to be relevant to the modified index. A further example with an $m=2$ Jacobi
form is given in \cite{LopesCardoso:1996zj}, where the computation of threshold corrections is through indirect methods
which we also will study in this paper and generalize to higher indexes. \cite{Kawai:1998md}, on the other hand, has some 
indications that Jacobi forms of several variables over root lattices of simple Lie algebras  may be relevant for heterotic
compactifications with Wilson lines. It explores the relation of such generalized Jacobi forms to Type IIA compactifications
on Calabi-Yau threefolds.

Turning on generic Wilson lines is important if one wishes to come up with detailed predictions for a possible Type II dual 
and the geometry of its compactification space, e.g. Gromov-Witten potentials of the compactifying 
Calabi-Yau manifold. So, we will continue along the way of \cite{LopesCardoso:1996nc},
\cite{LopesCardoso:1996zj} and extend their work by studying the relation of
Jacobi forms over general even lattices to heterotic compactifications 
with an arbitrary number of Wilson lines. The dependence of threshold corrections on Wilson lines has close relations to toroidal
compactifications and accordingly the presence of extended Jacobi forms in the context of $\mc{N} = 4$ compactifications
has already been noted in \cite{Kiritsis:1997hf}.

A useful property of Jacobi forms is that for a fixed weight and a fixed lattice they live in a finite dimensional
vector space.  Therefore, as in the context of modular forms or ordinary Jacobi forms, one can look at the 
first few Fourier coefficients of a quantity known to be a Jacobi form and use the finite dimensionality 
property to fix the whole function. Such arguments will be used on a number of examples which are connected in hypermultiplet moduli
space to an orbifold model. In this way, we will test our arguments with explicit computations using
exactly calculable orbifold conformal field theory.

Section 2 is an in depth analysis of the topics noted above. It starts with a general discussion of heterotic string
compactifications on $K3 \times T^2$ and associated threshold corrections. Section 2.1 discusses the overall structure of 
vector multiplet moduli space and explains in general terms why one would get a Jacobi form through the modified index computation.
In section 2.2, we show how the general discussion given in section 2.1 works in the context of orbifold compactifications.
Since our aim is to work with models connected to orbifold models on the hypermultiplet moduli space, in section 2.3 we study several
symmetry breaking patterns for $\mc{N} = 2$ gauge theories. In section 2.4, we start with an orbifold example and use the 
methods of section 2.3 to get a number of interesting models through Higgsing. Also, we show how the relevant Jacobi forms
can be obtained using
only partial information on the spectrum. These results are checked against explicit orbifold limit 
computations. After that, in section 2.5 we review the overall structure of threshold corrections and give explicit expressions
for the associated prepotential and gravitational coupling in terms of the Jacobi forms under study. Weyl chamber dependence
is another topic studied in this section. Lastly, in section 2.6 we study models with a single Wilson line at a generic point
of their vector multiplet moduli space. This is the simplest case of the formalism developed up to that point. The relevant Jacobi forms
are the Jacobi forms of \cite{Zagier}. For an index $m$ Jacobi form associated with such models threshold corrections
are given in terms of Siegel forms of the appropriate T-duality groups which turn out to be paramodular groups $\Gamma_m$. We give details for such models along the lines of \cite{LopesCardoso:1996nc},
\cite{LopesCardoso:1996zj}. We first review the cases $m=1$ and $m=2$, and then extend this for our examples with
$m=3$, $4$, and $5$.

Having discussed the details on the heterotic side, we turn our attention in section 3 to Type IIA compactifications 
and Calabi-Yau threefolds. One needs a dictionary to map vector multiplet moduli between those two cases. A possible way to accomplish
this for small $h^{1,1}$ values is to compare cubic prepotentials on both sides of the computation. A mapping without this
comparison is suggested in \cite{Kawai:1998md}. Our results on Weyl chambers in section 2 suggests a way to clarify some 
points and extend  the mapping to more general settings. In this form of the conjecture the moduli mapping
is unambiguously determined on the heterotic side.

Finally, relevant mathematical definitions, conventions and tools are gathered in the Appendix.

\section{\texorpdfstring{$\mathbf{\mathcal{N}=2}$}{N=2} Heterotic String Compactifications}

In this work, we will be interested in four dimensional string 
theories (on $\RR^{3,1}$) that can be obtained as compactifications of
six dimensional ($\RR^{5,1}$), $\mathcal{N}=(1,0)$ supersymmetric heterotic string models, on a torus $T^2$. 
The resultant four dimensional
theories have $\mathcal{N}=2$ spacetime supersymmetry and we will only study perturbative properties of these theories.

One way to obtain such a six dimensional theory is by geometrically compactifying ten dimensional ($\RR^{9,1}$)
heterotic string on a $K3$
surface. A $K3$ surface has a holonomy group restricted to $SU(2)$ and hence halves the number of supersymmetries of the ten
dimensional string theory. The massless spectrum of this six dimensional theory can be studied by starting with ten
dimensional, $\mathcal{N}=(1,0)$ supergravity coupled to $E_8 \times E_8$ or $Spin(32)/\ZZ_2$,
which are the low energy effective field theories of the two supersymmetric ten dimensional heterotic strings; and then 
dimensionally reducing these field theories on a $K3$ surface. 
Massless supermultiplets in the resulting six dimensional 
theory consist of
a gravity multiplet, a tensor multiplet, $N_v^6$ vector multiplets and $N_h^6$ hypermultiplets.
A tensor multiplet contains a single real scalar which is the heterotic dilaton. Hypermultiplets,
on the other hand, each contain four real scalars. Vector multiplets do not contain any scalars and their bosonic part 
consists only of a six dimensional vector field. One should also note that we restrict to 
perturbative models where there are no background five-branes. Five-brane backgrounds lead to 
additional tensor multiplets in the massless spectrum.

In such a scheme, one should also 
pick a non-trivial Yang-Mills background on the $K3$ surface for consistency. This can be easily seen by looking at the 
gauge invariant field strength of the $B$-field, $H = \diff B + \w_{3L} - \frac{1}{30} \w_{3Y}$, where the Chern-Simons terms
$\w_{3L}$ and $\w_{3Y}$ satisfy $\diff \w_{3L} = \tr R^2$ and $\diff \w_{3Y} = \Tr F^2$. In these equations, $R$ is the
curvature two-form, $F$ is the Lie algebra valued field strength two-form, and 
$\tr$ and $\Tr$ denote traces in the fundamental and adjoint representations, respectively. 
The background curvature is nontrivial for the $K3$ part and satisfies a topological constraint, 
\begin{equation}
 \int_{K3} \tr R^2 = 24.
\end{equation}
Therefore, if the $H$ field is well-defined,\footnote{Note that there is no $H$-flux on the $K3$
since we assume to have no background five-branes.} one needs to have
\begin{equation}
 \int_{K3} \diff H = \int_{K3} \tr R^2 - \frac{1}{30} \int_{K3} \Tr F^2 = 0,
\end{equation}
in other words, one needs to have a total of $24$ instantons turned on as a nontrivial gauge field background. 

Finally, one can compute the massless spectrum in six dimensions using anomaly cancellation and index theorems,
with respect to various ways of embedding the instantons in the $E_8 \times E_8$ or $Spin(32)/\ZZ_2$ gauge groups 
\cite{Green:1984bx}.

One can then form chains of such models where the subgroup in which the instantons are embedded
gets larger step-by-step. This leads to a cascade of models where the unbroken gauge group becomes smaller 
at each step. This can equivalently be described by charged hypermultiplet scalars getting a 
vacuum expectation value (vev) and breaking the gauge group.
Examples of such chains are discussed in \cite{Kachru:1995wm, Aldazabal:1995yw, Aldazabal:1996du}
and reviewed in \cite{Lust:1997kx}.

One such simple chain starts with $E_8 \times E_8$ heterotic string and embeds a $SU(2)$ bundle in the first $E_8$ with
$n_1$ instantons and a $SU(2)$ bundle in the second $E_8$ with $n_2$ instantons. It is common to parametrize these instanton
numbers by an integer, $k$, such that $(n_1, n_2) = (12+k, 12-k)$. For $k = 0,1,2$, this model has initially an
$E_7 \times E_7$ gauge symmetry and both $E_7$ factors can be broken through the chain 
\begin{equation}
 E_7 \rightarrow E_6 \rightarrow SO(10) \rightarrow SU(5) \rightarrow SU(4) \rightarrow SU(3) \rightarrow
 SU(2) \rightarrow 1.
\end{equation}
Throughout this chain, instantons are embedded in simple groups and the unbroken gauge group is the commutant
of this simple group in $E_8$. For larger $k$, it is not possible to completely 
break the gauge group in the $E_8$ factor with fewer
instantons since, at some point, the number of instantons becomes insufficient to support a background in a larger group.
This means for 
$k \geq 3$, one of the gauge groups can only be broken down to a terminal gauge group.

It is simple to find  the massless hypermultiplet spectrum of the $E_7 \times E_7$ theory from an index computation as
\begin{equation}
 \lp 4 + \frac{k}{2} \rp \rep{56,1} + \lp 4 - \frac{k}{2} \rp \rep{1,56} + 62 \rep{1,1},
\end{equation}
where we label irreducible $E_7$ representations by their dimensions.
Noting also the $266$ massless vector multiplets in the adjoint representations of $E_7$'s, one gets $N_h^6 - N_v^6 = 244$ as
required by the absence of gravitational anomalies for the $(1,0)$ supergravity in six dimensions 
with one tensor multiplet \cite{Lust:1997kx}.

More complicated examples can be obtained by turning on $U(1)$ backgrounds or semisimple backgrounds.
For example, going through the Higgsing process described above one can obtain the following massless 
hypermultiplet spectrum starting with the 
$U(1)$ background studied in \cite{Aldazabal:1996du} (where again $(12+k, 12-k)$ instantons are embedded in the $E_8$ factors):
\begin{equation}\label{eq:u1hyper}
 48 \rep{1}_q + 48 \rep{1}_{-q} + 149 \rep{1}_0,
\end{equation}
with the unbroken gauge group being $U(1)$.
Another example, \cite{Bershadsky:1996nh}, is obtained by embedding a $SU(3) \times SU(2)$ bundle in the first $E_8$ with
$(10, 4)$ instanton numbers and $10$ instantons
in the full $E_8$ for the second $E_8$. The hypermultiplet spectrum is then easily computed
as
\begin{equation}\label{eq:SU6model}
 99 \rep{1} + 10 \rep{6} + 10 \rep{\bar{6}} + 2 \rep{15} + 2 \rep{\overline{15}}   
\end{equation}
with respect to the unbroken gauge group $SU(6)$.
One can then break the gauge symmetry down to a $U(1)$ with
the same hypermultiplet spectrum as in \eqref{eq:u1hyper}.
We will describe the group theoretic details of such processes in the coming 
sections.

At this point, we start studying four dimensional theories that can be obtained as $T^2$ compactifications of the 
six dimensional theories
described above, where we will denote the final unbroken gauge group of the six dimensional theory as $G$.
The resulting theory has a gravity multiplet and a number of vector multiplets and hypermultiplets in its
massless spectrum. 

Four dimensional vector multiplets feature a complex scalar in addition to a four dimensional vector field in their bosonic part.
Hypermultiplets, on the other hand, have four real scalars, similar to six dimensions. Furthermore, hypermultiplets in 
six dimensions give hypermultiplets in the four dimensional theory.
 The gravity multiplet of six dimensions gives rise to a gravity multiplet in four dimensions as well as two vector multiplets. The complex
scalars of these vector multiplets, called $T$ and $U$ are moduli fields parameterizing the constant metric and constant 
$B$-field on $T^2$. The tensor multiplet gives another vector multiplet where its complex scalar, $S$, is the axio-dilaton
field. The weak coupling limit of the heterotic string 
is $\Im S = 4 \pi / g_s \rightarrow \infty$.
Lastly, vector multiplets in six dimensions give rise to vector multiplets in four dimensions.

The potential
associated with the vector multiplet scalars remain zero if commuting members of vector multiplet scalars gain a vev 
(which is equivalent to turning on
Wilson lines on the torus $T^2$). Therefore, at a generic point in the moduli space of vector multiplets, scalars in the 
Cartan subalgebra of the gauge group gain a vev and every state that is charged gains mass. This leaves the gravity 
multiplet, $N_h$ neutral fields in the hypermultiplet spectrum and $N_v = 3 + \rank(G)$ 
vector multiplets as massless fields. The number
three comes from the vector multiplets carrying the scalars $S$, $T$ and $U$. In the following work, we will call the scalars
of the $s \equiv \rank(G)$ vector multiplets  Wilson lines and denote them by $V_i$ where $i = 1,..., s$, although, in general,
those moduli will be some combinations of $T^2$ moduli and Wilson lines on $T^2$.

An interesting physical quantity one can compute in such theories is the modified index defined in \cite{Cecotti:1992qh}.
We will compute this index in the form
\begin{equation}
 \I = - \frac{ \ii}{\n^2} \text{Tr} \lp  J_0 (-1)^{J_0} q^{L_0 - c/24} \bar{q}^{\bar{L}_0 - \bar{c}/24} \rp,
\end{equation}
where the trace is taken over the Ramond sector of the internal CFT and $J_0$ is the zero mode of the $U(1)$ current 
which lies in the $N =2$
superconformal symmetry of the internal CFT. We will adopt the convention that $N$ denotes worldsheet supersymmetry; 
whereas $\mc{N}$ denotes spacetime supersymmetry.

Our main interest in this quantity lies in the fact that
its integral over the fundamental domain of $SL(2,\ZZ)$ in the form 
\begin{equation}\label{eq:gravthre}
 \Delta_\text{grav} = \int_\mc{F} \frac{\diff^2 \T}{\T_2} \lb 
    - \frac{ \ii}{\n^2} \text{Tr} \lp  J_0 (-1)^{J_0} q^{L_0 - c/24} \bar{q}^{\bar{L}_0 - \bar{c}/24} \rp 
    \lp E_2(q) - \frac{3}{\pi \T_2}  \rp  - b_\text{grav} \rb 
\end{equation}
and
\begin{equation}\label{eq:gaugethre}
 \Delta_\text{gauge} = \int_\mc{F} \frac{\diff^2 \T}{\T_2} \lb 
    - \frac{ \ii}{\n^2} \text{Tr} \lp  J_0 (-1)^{J_0} q^{L_0 - c/24} \bar{q}^{\bar{L}_0 - \bar{c}/24} 
    \lp Q^2 - \frac{1}{8 \pi \T_2} \rp   \rp 
     - b_\text{gauge}	 \rb 
\end{equation}
gives gravitational and gauge coupling threshold corrections, respectively \cite{Antoniadis:1992rq}.
In these expressions, $E_2$ is the order two Eisenstein series, 
$\T_2 = \Im \T$ and $Q$ is a generator of the gauge group for which we are computing the threshold correction. We will be more
explicit about our conventions in the coming sections.

The theories we consider here have $\mathcal{N}=(1,0)$ supersymmetry in six dimensions. This means that the internal CFT
for the four dimensional theory is the sum of a $N=(0,2)$ supersymmetric free theory with central charge $(2,3)$ corresponding to
the torus and $N=(0,4)$ supersymmetric CFT with central charge $(20,9)$ \cite{Banks:1988yz}. 
Following the discussion
in the section 3 of \cite{Harvey:1995fq}, one sees that the trace involved in $\I$ reduces to the Witten index for the 
$N=(0,4)$ supersymmetric part and hence it is invariant under its smooth deformations. Hypermultiplet moduli of the four
dimensional theory comes purely from this factor and hence $\I$ is invariant under smooth changes of hypermultiplet moduli.

This is, of course, different for vector multiplet moduli. The scalars of vector multiplets have vertex operators of the form
$\delbar X^\pm (\bar{z}) J^a(z)$ where $X^\pm$ are the free scalars of the $(c,\bar{c})=(2,3)$ factor and $J^a$ is a gauge
current. Finding the dependence of threshold corrections on vector multiplet moduli for the models
we consider is one of the goals of this paper.
The hypermultiplet moduli independence of the modified index allows its computation for a wide class of theories which are
smoothly connected to orbifold limits where one can explicitly compute the modified index and its vector multiplet moduli 
dependence. This is essentially done in \cite{Henningson:1996jz} where the modified index is expressed in terms of
several lattice sums. 
Our main aim will be to emphasize the modular properties of the modified index, give its relations
to Jacobi forms (of several variables) and display the power of these modular restrictions on several examples and to check
these results with orbifold computations.

\subsection{Vector Multiplet Moduli Space and Jacobi Forms from Modified Index}
The classical moduli spaces of vector multiplet moduli for $\mc{N} = 2$, $D=4$ supergravity theories are strictly restricted. Peccei-Quinn
symmetry requires them to be a product of two factors where one is parametrized by the axio-dilaton field alone. Then, it is a
theorem in supergravity \cite{Ferrara:1989py} that the only such product manifold is of the form
\begin{equation}
 \frac{SU(1,1)}{U(1)} \otimes \frac{SO(n,2)}{SO(n) \otimes SO(2)}.
\end{equation}
In the models we consider, where one compactifies a six dimensional theory on a two-torus, moduli that spans the second factor
are the torus moduli $T$, $U$ and Wilson lines $V^i$ so that $n=s+2$. This picture is further refined in string theory
by discrete identifications
on the moduli space. The identifications on the ${SO(n,2)}/{SO(n) \otimes SO(2)}$ part can be obtained by the action of a
T-duality group. We will discuss this aspect later in our discussion. Particular examples of such duality groups
include paramodular groups.

Now, it is necessary to describe the details of the final, unbroken gauge group of the six dimensional theory before 
compactification. (This is equivalent to describing the $4D$ classical gauge group enhancement
one would observe when the Wilson lines on the torus
are switched off.) 
Let the gauge group be a product of simple factors coming from simply laced Lie algebras and a number of $U(1)$'s. 
\begin{equation}\label{eq:group}
 G = \prod G_i \times \lb  U(1) \rb^k.
\end{equation}
We will only deal with the case where the simple factors are generated at level 1.
Then, the current algebra corresponding to the Cartan subalgebra of the simple factors together with the $U(1)$ 
factors can be factored out by compact scalars on a $s$-dimensional torus (Frenkel-Kac-Segal construction, 
\cite{FrenkelKac, Segal}).
This is the torus generated by exponentiating this maximally commuting part of the
total gauge Lie algebra \cite{Goddard:1986bp}.
In other words, we take $s$ left moving scalars $X^i$ normalized so that 
$\del X^i(z) \del X^j(w) = \delta^{ij} / (z-w)^2 + \ldots$
$U(1)$ generators are $\vec{j}(z) = \ii \del \vec{X}(z)$ and we identify points as
\begin{equation}
 \vec{X} \sim \vec{X} + 2 \pi \lt,
\end{equation}
where the lattice, $\lt$, defines the torus, $\RR^s/(2 \pi \lt) $, described above. The usual Euclidean metric induced
by the $\del X \del X$ OPE gives a bilinear form on this lattice.

The lattice, $\lt$, is a direct sum of the simple factors' root lattices and a number of one-dimensional lattices 
corresponding to $U(1)$ 
factors, where each of these lattice factors are orthogonal to each other.  
\begin{equation}
 \lt = \lp \oplus \lt_{j,R} \rp \oplus \lp \oplus_{i=1}^k \langle m_i \rangle \rp,
\end{equation}
where $\lt_{i,R}$ denotes the root lattice corresponding to $G_i$, and $\langle m \rangle$ is a one dimensional lattice, 
$\ZZ \vec{\beta}$, with $\vec{\beta} . \vec{\beta} = m$. The charges with respect to those $U(1)$'s lie on the dual lattices
$\langle m_i \rangle^*$. Finally, we require $m_i$'s to be even integers. 
The fact that the lattice $\lt$ is integral and that the
charge lattice lies in the dual lattice $\lt^*$ is what makes the theory invariant under a 'spectral flow'. 

Next, we take an integral basis to the lattice $\lt$ (for definiteness we will take the simple roots of the simple factors
to be in this basis) and denote the members of this basis by $\vec{\beta_i}$.\footnote{In the following text, 
when we talk about a $U(1)$ 
generated by or in the 
direction of a vector $\vec{\beta} \in \lt$, we mean the current $\vec{\beta}.\vec{j}(z)$. } $\vec{\beta_i}$'s induce a lattice
metric as $d_{ij} = \vec{\beta_i} . \vec{\beta_j}$. Further, we can define a dual basis for $\lt^*$ by
$\vec{\gamma^i} = d^{ij} \vec{\beta_j}$, where $d^{ij}$ is the inverse of the lattice metric. 

We can now separate the contribution of the gauge currents in the holomorphic part of the energy tensor as
\begin{equation}
 T(z) = \frac{1}{2} \lp \vec{j}(z) \rp^2 + T'(z),
\end{equation}
where $T'(z)$ is independent of the scalars $\vec{X}$.

Then, we take a generic state that has charge $\vec{q} \in \lt^*$ and is generated by a primary field 
$\Sigma(z, \bar{z})$ in the 
$(c, \bar{c}) = (20, 9)$ internal CFT plus some other components coming from the non-compact six dimensional part.
Then, we separate the contribution from the charge as 
\begin{equation}
 \Sigma(z, \bar{z}) = : e^{\ii \vec{q} . \vec{X}}(z): W(z, \bar{z}).
\end{equation}
$W(z, \bar{z})$ denotes contributions independent of $\vec{X}$ (up to oscillator contributions
coming from polynomials in $\del \vec{X}$)
and we denote its $L_0$ ($\bar{L}_0$) eigenvalues by $h_L$ ($h_R$).
It follows that
\begin{equation}
 \ii \del \vec{X}(z) \Sigma(w, \bar{w}) = \frac{\vec{q}}{z-w} \Sigma(w, \bar{w}) + \ldots
\end{equation}
or, expressed in another way,
\begin{equation}
 \vec{j}_0 | \Sigma \rangle = \vec{q} | \Sigma \rangle,
\end{equation}
where $| \Sigma \rangle$ is the state generated by $\Sigma(w, \bar{w})$.
Also, we can see that the contribution of the charge part to $L_0$ eigenvalue is $\vec{q}.\vec{q} /2$.
If one decomposes $\vec{q}$ as $\vec{q} = k_i \vec{\gamma^i}$ with $(k_i) \in \ZZ^s$, this contribution is 
equivalently $k_i d^{ij} k_j / 2$.

We now consider vertex operators
$V_{\vec{\lambda}}(z) = :e^{\ii \vec{\lambda} . \vec{X}}(z):$ for some $\vec{\lambda} \in \lt$. 
This is a well defined operator because
$\lt \subset \lt^*$. 
Also, since $V_{\vec{\lambda}}(z) V_{\vec{q}}(w) \sim (z-w)^{\vec{\lambda} . \vec{q}} V_{\vec{\lambda}+\vec{q}}(w) + \ldots$,
this operator is local with respect to every state. That means there is a bijective pairing between states of charge $\vec{q}$
and $\vec{\lambda} + \vec{q}$ where 
$W(z, \bar{z})$ contributions are matched.

Now, we compactify on a two-torus with arbitrary constant metric and B-field. Then we turn on Wilson lines via deformations
of the string Lagrangian with a term of the form
\begin{equation}
 \epsilon^{\alpha \beta} \del_\alpha X^\mu \vec{A}_\mu . \del_\beta \vec{X},
\end{equation}
where $X^\mu$ with $\mu = \pm$ are coordinates on the torus.

One can now go through the same procedure as we went through in the six dimensional theory. There are now $s+4$
$U(1)$ currents and one can write currents in terms of $s+4$ compact scalars. The lattice that determines the torus is
constructed by rotating $U \otimes U \otimes \lt$ by a member of $SO(s+2, 2)$ where $U$ is the standard hyperbolic lattice and
the member of $SO(s+2, 2)$ that rotates the standard lattice is determined by moduli.

Very schematically, $\vec{j}_0 |m_i, n_i, k_i \rangle = \vec{p}_L |m_i, n_i, k_i \rangle $,
$\vec{\bar{j}}_0 |m_i, n_i, k_i \rangle = \vec{p}_R |m_i, n_i, k_i \rangle $ and the contribution of these scalars
to the energy tensor is $\bar{T}_X (\bar{z}) = \frac{1}{2} :\vec{\bar{j}}.\vec{\bar{j}}(\bar{z}):$ and 
$T_X (z) = \frac{1}{2} :\vec{j}.\vec{j} (z):$. Through the parametrization of $SO(s+2, 2)/SO(s+2) \otimes SO(2)$
by a vector $y = (T,U,\vec{y}) \in \CC^{17,1}$, as used in \cite{Harvey:1995fq}, 
one gets
\begin{equation}\label{eq:pL}
 \frac{p_L^2 - p_R^2}{2} = \frac{1}{2} \vec{q}.\vec{q} - m_1 n_1 + m_2 n_2,
\end{equation}
and
\begin{equation}\label{eq:pR}
 \frac{p_R^2}{2} = \frac{1}{-2(y_2,y_2)} 
      \left|  \vec{q} . \vec{y} + m_1 U + n_1 T + m_2 - n_2 \frac{(y,y)}{2}\right|^2.
\end{equation}
The inner product $(y,y')$ appearing above is defined by 
$(y,y')  = -T U' - U T' + \vec{y} . \vec{y'}$.
Using (integral) charges $k_i$ defined by $\vec{q} = k_i \vec{\gamma^i}$ and
parameterizing Wilson lines by $\vec{y} = V^i \vec{\beta_i} $ these expressions can be given in terms of the 
$\lt$-lattice metric $d_{i j}$ as
\begin{equation}
 \frac{p_L^2 - p_R^2}{2} = \frac{1}{2} k_i d^{i j} k_j - m_1 n_1 + m_2 n_2,
\end{equation}
and
\begin{equation}
 \frac{p_R^2}{2} = \frac{1}{-2(y_2,y_2)} 
      \left|  k_i V^i + m_1 U + n_1 T + m_2 - n_2 \frac{(y,y)}{2}\right|^2,
\end{equation}
where
\begin{equation}
 (y,y) \equiv  y^a \eta_{a b} {y}^b = - 2 T U + V^i d_{i j} V^j .
\end{equation}

Following a similar argument to the one above we see that
states with charge $(m_i, n_i, k_i) \in \lp U \oplus U \oplus \lt^* \rp$ are bijectively matched
with respect to their $W(z, \bar{z})$ contributions
when the charge is shifted
by a member of its dual lattice, $\G \equiv U \oplus U \oplus \lt$ (which is the lattice that creates the torus
before the moduli dependent rotation).

Using this matching, it is possible to factorize the vector multiplet moduli dependence in a number of interesting physical quantities
such as the modified index or the partition function. Suppose we would like to compute a quantity which involves a trace
over the internal CFT. Furthermore, suppose that only insertion in the trace that
involves $(X^\pm, \vec{X})$ part of the CFT is $q^{L^X_0 - c^X/24} \bar{q}^{\bar{L}^X_0 - \bar{c}^X/24}$.
In this case
\begin{align}
 \Tr_{\mc{H}_\text{int}} \lb q^{L^X_0 - c^X/24} \bar{q}^{\bar{L}^X_0 - \bar{c}^X/24} \ldots \rb 
&= \sum_{(m,n,k) \in \G^*} q^{p_L^2/2} \bar{q}^{p_R^2/2} \Tr_{\mc{H}^{m,n,k}_\text{int}} \lb \ldots \rb    \\
    &= \sum_{\mu \in \lt^* / \lt} \lp \sum_{(m,n,k-\mu) \in \G} q^{p_L^2/2} \bar{q}^{p_R^2/2} \rp  
				  \lp \Tr_{\mc{H}^{0,0,\mu}_\text{int}} \lb \ldots \rb \rp \\
    &= \sum_{\mu \in \lt^* / \lt} Z_{\G,\mu}(\T, T, U, V^i) h_\mu  \label{eq:thetadcmp},
\end{align}
where $\mc{H}^{m,n,k}_\text{int}$ is the vector space of states with charge $(m_i, n_i, k_i)$,
\begin{equation}
  Z_{\G,\mu}(\T, T, U, V^i) \equiv \sum_{(m,n,k-\mu) \in \G} q^{p_L^2/2} \bar{q}^{p_R^2/2} ,
\end{equation}
and
$ h_\mu \equiv \Tr_{\mc{H}^{0,0,\mu}_\text{int}} \lb \ldots \rb$.

For the modified index of the theories we consider,
this leads to an interesting result. $ h_\mu$ is independent of hypermultiplet moduli and $\bar{q}$, and hence
we write $h_\mu = h_\mu (q)$. Furthermore, we note that $\T_2 \I (E_2 - 3 / \pi \T_2)$ should be modular invariant for 
the integral, \eqref{eq:gravthre}, to make sense.\footnote{Note that $b_{\text{grav}}$ term in the integral is an IR regulator.} Of course, this should
be correct for any value of vector multiplet moduli and in particular for $V^i = 0$. For $V^i = 0$,
$Z_{\G,\mu}$ factorizes as
\begin{equation}
 Z_{\G,\mu}\Big|_{V^i = 0} = Z_{2,2} \tht_{\lt, \mu}(\T, \vec{z}=0),
\end{equation}
where $\tht_{\lt, \mu}$ is a theta function associated to $\lt$,
\begin{equation}
 \tht_{\lt, \mu_i}(\T, z^i)
      = \sum_{k_i \equiv \mu_i (\text{mod } d_{i j} n^j)} q^{k_i d^{i j} k_j / 2} y_1^{k_1} \ldots y_s^{k_s},
      \text{ where } y_i \equiv e^{2 \pi \ii z^i},
\end{equation}
and $Z_{2,2}$ is a sum over torus momentum and winding numbers.

Since $ Z_{2,2} \T_2 (E_2 - 3 / \pi \T_2)$ transforms under modular transformations with weight $2$, the factor
\begin{equation}
 \sum_{\mu \in \lt^* / \lt} \tht_{\lt, \mu}(\T, \vec{z}=0) h_\mu(\T)
\end{equation}
transforms with weight $-2$. This finally implies that $\phi_{-2, \lt}(\T, \z)$ defined by
\begin{equation}
 \phi_{-2,\lt}(\T, \z) = \sum_{\mu \in \lt^* / \lt} \tht_{\lt, \mu}(\T, \vec{z}) h_\mu(\T)
\end{equation}
is a weight $-2$ Jacobi form with respect to the lattice $\lt$.\footnote{See the Appendix for the definition of Jacobi forms.}

More explicitly,
\begin{equation}
 \phi_{-2,\lt}(\T, \z) = \sum_{n, k_i} c(n,k_i) q^n y_1^{k_1} \ldots y_s^{k_s}.
\end{equation}
Here, the variables $y_s$ counts charges in the Cartan subalgebra of $G$.

By the elliptic transformation property of Jacobi forms, the Fourier coefficients $c(n,k_i)$ depend only on
$\Delta \equiv n - k_i d^{i j} k_j / 2$ and $k_i(\text{mod } d_{i j} n^j)$, which is obvious from the theta decomposition
form we have above. For the modified index, $\Delta = N_L + h_L -1$, where $N_L$ is the total left oscillator number and
$h_L$ is the contribution of $(c, \bar{c}) = (20, 9)$ internal CFT to $L_0$ excluding the contribution from the charge part.
This means that  $c(n,k_i)$ vanishes unless
$\Delta = n - k_i d^{i j} k_j / 2 \geq -1$. In particular, the only $q^{-1}$ term comes from the unphysical tachyon
\cite{Harvey:1995fq} as 
\begin{equation}
 \phi_{-2,\lt}(\T, \z) = -\frac{2}{q} + O(1), \text{ as } \Im \T \rightarrow + \infty.
\end{equation}
Moreover, multiplying $\phi_{-2,\lt}$ by the cusp form $\Delta(q) = \n^{24}$ one gets a weight $10$, holomorphic Jacobi form
associated with the lattice $\lt$. Also, trivially extending the discussion in \cite{Harvey:1995fq}, we see that we can
interpret $\phi_{-2,\lt}(\T, \z)$ as
\begin{equation}
 Z_{2,2} \phi_{-2,\lt}(\T, \z) = 2 \lp  \sum_{\text{Hyper BPS}} q^\Delta y_1^{k_1} \ldots y_s^{k_s}
				  - \sum_{\text{Vector BPS}} q^\Delta y_1^{k_1} \ldots y_s^{k_s} \rp,
\end{equation}
where to be exactly correct one should set the Wilson lines to zero for the 
$\G$ momentum contribution to $L_0$ and ignore constraints coming from level matching. Also,
note that the terms vector BPS and hyper BPS multiplets in this expression
are used in the sense of \cite{Harvey:1995fq}, where one separates them according to their structure in the right-moving part
of the internal CFT. One important point to note is that the gravity multiplet contributes to this sum as a 	
vector BPS multiplet $q^0$ term with zero charge.

Finally, let us make a few comments on T-duality. The norm and conjugacy class preserving
automorphisms of the lattice $\G$ for the norm
\begin{equation}
 p_L^2 - p_R^2 = k_i d^{i j} k_j - 2 m_1 n_1 + 2 m_2 n_2,
\end{equation}
can be compensated by a corresponding transformation of the moduli $(T, U, V^i)$ so that $p_R^2$ and hence the spectrum is also
preserved. These transformations form  part of the T-duality group transforming vector multiplet moduli $T$, $U$ and $V^i$. 
If there are identifications 
between $\mc{H}^{0,0,\mu}_\text{int}$, then it is also possible that the
T-duality group can be extended by lattice automorphisms mixing 
some of the conjugacy classes.
Notice that the form of $\Delta_{\text{grav}}$ makes it manifestly T-duality invariant.

In \cite{Neumann:1996is},  it is shown that for $\lt  = \langle 2m \rangle$ where $m \in \ZZ^+$, the
T-duality transformations
described above generate the extended paramodular group $\G_m^+$. We will describe this group later in more detail.
We will generically denote such discrete transformations by $O(\G)$. Then, 
the perturbative moduli space will have the following component
from $(T, U, V^i)$ moduli:
\begin{equation}
 O(\G) /  \frac{SO(n,2)}{SO(n) \otimes SO(2)}.
\end{equation}

One should also note that a similar argument applies for the trace expression generating $24 \Delta_{\text{gauge}} - 
\Delta_{\text{grav}}$,
\begin{equation}
 \B \equiv - \frac{ \ii}{\n^2} \text{Tr} \lb  J_0 (-1)^{J_0} q^{L_0 - c/24} \bar{q}^{\bar{L}_0 - \bar{c}/24} 
	\lp  24 Q^2 - E_2 \rp		\rb,
\end{equation}
provided that $Q$ is a generator  of a gauge group for which no Wilson lines
are turned on. That may happen if, for example, the gauge group of the six-dimensional theory is, say,
$G \times SU(2)$ but Wilson lines are switched on only for $G$.\footnote{Remember that when we say Wilson line we actually
mean a combination of Wilson lines and torus moduli in the form $V^i$.} Repeating the argument above, one gets a factorization
\begin{equation}
 \B = \sum_{\mu \in \lt^* / \lt} Z_{\G,\mu}(\T, T, U, V^i) f_\mu.
\end{equation}
Taking $V_i$ to zero and looking at the modular properties of $\B$ one can deduce that $f_\mu$'s combine into a 
weight $0$ Jacobi form for the lattice $\lt$, which we will denote by $\psi_{0,\lt} (\T, \z)$.

The requirements we give for $G$ are satisfied for geometric compactifications of the ten dimensional heterotic string
theories where the unbroken gauge group at the six-dimensional level are created by a suitable part of the 
original, level one, $E_8 \times E_8$ or $Spin(32)/\ZZ_2$ current algebra at ten dimensions.
We will show in the next section how this factorization also works for the modified index computed at an orbifold point
more explicitly.

Finally, 
before leaving this section, note that the 'spectral flow' property described in this section
is in the spirit of \cite{Kraus:2006nb}, where, in comparison, we have the charge lattice more explicit here, and it 
is very similar to
the 'spectral flow' described in \cite{deBoer:2006vg, Gaiotto:2006wm} for M-theory and Type-II compactifications
on a Calabi-Yau threefold.

\subsection{Modified Index at Orbifold Limits}
In this section, we will be interested in $\mc{N} = 2$ orbifolds of 
$\G^{16} = E_8 \times E_8'$ or $\G^{16} = Spin(32)/\ZZ_2 $ heterotic string. We start by compactifying 
the heterotic string on $T^4 \times T^2$. Then, we introduce complex coordinates on $T^4$ and assume that it admits a $\mathbb{Z}_N$
symmetry,
$z_1 \rightarrow \exp(2 \pi i a / N) z_1$ and $z_2 \rightarrow \exp(-2 \pi i a / N) z_2$ where
$a=0,1,\cdots,N-1$. Such tori exist for $N=2,3,4,6$. We will now mod out by that $\mathbb{Z}_N$ symmetry. Since
there are fixed points of those $\mathbb{Z}_N$ transformations; 
although there is no curvature away from the fixed points, there will be curvature concentrated on them since
there is a conical singularity with a monodromy contained in $\mathbb{Z}_N$. These
monodromies can be included in an $SU(2)$ group and in this way we get an orbifold limit of $K3$ surfaces. 

To preserve modularity (or equivalently to cancel spacetime anomalies) $\mathbb{Z}_N$ should also act on the
gauge bundle over the orbifold. We will be concerned with the case in which we are going to implement this action via a shift, 
$\frac{a}{N} \vec{\g}$, on the $\G^{16}$ lattice. Here $\vec{\g} \in \G^{16}$ and $a \in \ZZ$
to have an action included in $\mathbb{Z}_N$. Note that
none of the $U(1)$ generators of the original string theory are projected out of the spectrum. However, a root, $\vec{r}$,
survives only if
$N | \vec{r}.\vec{\g}$. Here we are using the natural Euclidean metric induced on $\G^{16} \otimes \RR$.

Next, we vary the constant metric and B field on $T^2$ as well as Wilson lines on $T^2$ giving a vev to 
 scalars in the Cartan subalgebra (CSA)as in the last section. Coordinates on vector multiplet moduli space can be described by a vector, 
$y = (T, U, \vec{y} ) \in \CC^{17,1}$ as in the last section.

Then, as given in \cite{Henningson:1996jz} we have explicitly
\begin{equation}
 \mc{I} = -\frac{1}{4 N \n^2} \sum_{a,b=1}^{N} Z_{a,b}^{\text{K3}} Z_{a,b}^{\text{torus}},
\end{equation}
where
\begin{equation}
 Z_{a,b}^{\text{K3}} = k_{a b} q^{-(a/N)^2} \n^2 \tht_1^{-2} \lp \T \frac{a}{N} + \frac{b}{N} \Big| \T \rp,
\end{equation}
and
\begin{equation}
 Z_{a,b}^{\text{torus}} =  e^{-2 \pi \ii \frac{ab}{N^2} \vec{\g}^2} \n^{-18} 
	      \sum_{\vec{R} \in \G^{16} + \frac{a}{N} \vec{\g}, m, n} e^{2 \pi \ii \frac{b}{N} \vec{R} . \vec{\g}}
			      q^{\tilde{p}_{a L}^2 /2} \bar{q}^{\tilde{p}_{a R}^2 /2}.
\end{equation}
Above $k_{a b}$ are constants determined by modularity. They are given in \cite{Henningson:1996jz} and \cite{Stieberger:1998yi}.
$\tilde{p}_{a L}^2$ and $\tilde{p}_{a R}^2$ are defined as in equations \eqref{eq:pL} and \eqref{eq:pR} where we use
$\vec{R}$ instead of $\vec{q}$.

When vector multiplet moduli are turned off, the gauge group of an orbifold model is as in \eqref{eq:group}.
Suppose a vector $\vec{\beta}' \in \G^{16}$ creates one of those $U(1)$ factors and it is a primitive vector of $\G^{16}$. 
Charges with respect to this vector then can be fractional, as there are states in the twisted sector 
surviving the orbifold projection such that their charges are multiples of $\text{gcd} \lp \vec{\beta}' . \vec{\g}, N \rp / N$. So, to make 
the charges lie on the dual of the lattice of the compact scalar generating this $U(1)$, one should take 
$\ZZ \vec{\beta} \equiv \ZZ \lp \frac{N}{\text{gcd} \lp \vec{\beta}' . \vec{\g}, N \rp} \vec{\beta}' \rp$ as the factor 
appearing in $\lt'$. Here, $\lt'$ is the lattice of compact scalars generating the gauge current.
This lattice now satisfies the hypothesis given in the previous section.

Suppose, now, we move in the hypermultiplet moduli space away form the orbifold point by Higgsing some parts of the gauge group (giving a
vev to charged hypermultiplets). Out of the compact scalars of the initial orbifold, only some linear combinations can now 
be used to turn on Wilson lines. The directions in which Wilson lines can be turned on should be orthogonal to the 
charge vectors of hypermultiplets getting a vev. This determines a $s$-complex dimensional subspace, $Y^s$ of $\CC^{16}$.
This means that the final unbroken gauge group will be generated by
compact scalars living in a $s$-dimensional torus generated by $\lt \equiv Y^s \cap \lt' \subset \lt'$.\footnote{
We will break the gauge symmetry by giving a vev to hypermultiplet pairs with charge $\pm \lambda \in \lt'^*$ as will
be described in the next section. It is not hard to prove inductively that with such a breaking pattern, the lattice $\lt$ has
rank $s$.}

By the construction
of $\lt'$ we described above, $\lt$ is an even lattice, with charges lying in the dual lattice $\lt^*$. Moreover, if
$\vec{\beta_i}$ is an integral basis for $\lt$ (where we again use the root vectors for the simple group contributions)
one has $\vec{\beta_i} . \vec{\g} = s_i N$ for some integers $s_i$. This will be crucial in displaying the factorization
for the modified index.

By the hypermultiplet moduli independence, we can compute the modified index for this final model using the orbifold limit expression,
where now, $\vec{y} = V^i \vec{\beta_i}$ and we define the lattice metric $d_{i j}$ and the dual basis
$\{\vec{\g^j} \} $ as before.
\begin{equation}
 \I = -\frac{1}{4 N \n^{18}} \sum_{a,b=1}^{N}  k_{a b} q^{-(a/N)^2} \tht_1^{-2} \lp \T \frac{a}{N} + \frac{b}{N} \Big| \T \rp
      \sum_{\vec{r} \in \G^{16}, m, n} e^{2 \pi \ii \frac{b}{N} \vec{r} . \vec{\g}}
			      q^{\tilde{p}_{a L}^2 /2} \bar{q}^{\tilde{p}_{a R}^2 /2}.
\end{equation}
Let us partition $\G^{16}$ into disjoint sets according to their contribution to charges with respect to $\vec{\beta_i}$.
We define
\begin{equation}
 S_{k_i} \equiv \{  \vec{r} \in \G^{16} | \vec{\beta_i}.\vec{r} = k_i \text{ for } i=1,\ldots,s \}.
\end{equation}
The crucial observation is that for any set of integers $n^1, \ldots, n^s$ one has
\begin{equation}\label{eq:shift}
 S_{k_i + d_{i j} n^j} = S_{k_i} + n^j \vec{\beta_j},
\end{equation}
reminiscent of the spectral flow property described in the previous section.
Further, for any $\vec{r} \in S_{k_i - a s_i}$ one has
\begin{align}
 \vec{R}_a &= \vec{r} + \frac{a}{N} \vec{\g} = \lp \vec{r} + \frac{a}{N} \vec{\g} - k_i d^{i j} \vec{\beta_j} \rp 
	+ k_i d^{i j} \vec{\beta_j}  \\
    &=  \lp \vec{r} + \frac{a}{N} \vec{\g} - k_i \vec{\g^i} \rp 
	+ k_i  \vec{\g^i}.
\end{align}
Note that the second factor is in the lattice $\lt^*$ and the first factor is perpendicular to the space $\lt \otimes \RR$.
So, in comparison to equations \eqref{eq:pL} and \eqref{eq:pR}, we define $p_L^2$ and $p_R^2$ via:
\begin{align}
 \frac{\tilde{p}_{a L}^2 - \tilde{p}_{a R}^2}{2} &=
    \frac{1}{2}   \lp \vec{r} + \frac{a}{N} \vec{\g} - k_i \vec{\g^i} \rp^2 + \frac{1}{2} k_i d^{i j} k_j - m_1 n_1 + m_2 n_2, \\
	&= \frac{1}{2}  \lp \vec{r} + \frac{a}{N} \vec{\g} - k_i \vec{\g^i} \rp^2 + \frac{p_L^2 - p_R^2}{2}, 
\end{align}
and
\begin{equation}
 \frac{\tilde{p}_{a R}^2}{2}  =  
 \frac{\left| k_i V^i + m_1 U + n_1 T + m_2 + n_2 \lp T U -\frac{1}{2}V^i d_{i j} V^j \rp \right|^2}{4 \lp T_2 U_2 - \frac{1}{2}V_2^i d_{i j} V_2^j \rp} 
 = \frac{p_R^2}{2}.
\end{equation}

For $f_{a b} (\T)$ defined as
\begin{equation}
f_{a b} (\T) = - \frac{1}{4 N \n^{18}} k_{a b} q^{-(a/N)^2} \tht_1^{-2} \lp \T \frac{a}{N} + \frac{b}{N} \Big| \T \rp,
\end{equation}
we get
\begin{equation}
 \I = \sum_{a,b=1}^{N} f_{a b} \sum_{m_i,n_i} \sum_{k_i} q^{p_L^2/2} \bar{q}^{p_R^2/2} 
	\sum_{\vec{r} \in S_{k_i - a s_i}} e^{2 \pi \ii \frac{b}{N} \vec{r} . \vec{\g}}
			      q^{ \lp \vec{r} + \frac{a}{N} \vec{\g} - k_i \vec{\g^i} \rp^2 / 2} .   
\end{equation}

Now, we separate the sum over $(k_i) \in \ZZ^s$ as a sum over $k_i \equiv \mu_i  (\text{mod } d_{i j} n^j)$ and a sum over
$  \mu_i  (\text{mod } d_{i j} n^j) $. Noting the following relation, using \eqref{eq:shift} and that $\vec{\beta_i} . \vec{\g} = s_i N$
\begin{equation}
 \sum_{\vec{r} \in S_{k_i - a s_i}} e^{2 \pi \ii \frac{b}{N} \vec{r} . \vec{\g}}
			      q^{ \lp \vec{r} + \frac{a}{N} \vec{\g} - k_i \vec{\g^i} \rp^2 / 2} = 
	\sum_{\vec{r} \in S_{\mu_i - a s_i}} e^{2 \pi \ii \frac{b}{N} \vec{r} . \vec{\g}}
			      q^{ \lp \vec{r} + \frac{a}{N} \vec{\g} - \mu_i \vec{\g^i} \rp^2 / 2},
\end{equation}
we find
\begin{equation}
 \I = \sum_{\mu_i  (\text{mod } d_{i j} n^j)} \lp  
 \sum_{\substack{k_i \equiv \mu_i  (\text{mod } d_{i j} n^j), \\ m_i,n_i}}
    q^{p_L^2/2} \bar{q}^{p_R^2/2} \rp
    \lp
	\sum_{\substack{\vec{r} \in S_{\mu_i - a s_i},\\ a,b}}  e^{2 \pi \ii \frac{b}{N} \vec{r} . \vec{\g}}
			      q^{ \lp \vec{r} + \frac{a}{N} \vec{\g} - \mu_i \vec{\g^i} \rp^2 / 2}f_{a b} \rp  . 
\end{equation}
This is of the form \eqref{eq:thetadcmp} as claimed, where 
\begin{equation}
 h_\mu (\T) = \sum_{\substack{\vec{r} \in S_{\mu_i - a s_i},\\ a,b}}  e^{2 \pi \ii \frac{b}{N} \vec{r} . \vec{\g}}
			      q^{ \lp \vec{r} + \frac{a}{N} \vec{\g} - \mu_i \vec{\g^i} \rp^2 / 2} f_{a b}.
\end{equation}

Also, note that we still have the factorization above if there is an insertion of the form $(\vec{R} . \vec{Q})^2$ in the sum
provided that $\vec{Q}.\vec{\beta_i} = 0$ for all $i$.\footnote{The normalization of $Q^2$ can be deduced from
\cite{Henningson:1996jz} as $Q^2 = \frac{(\vec{P}.\vec{Q})^2}{2(\vec{Q}.\vec{Q}) }$ for a state with charge
$\vec{P}$ with respect to $\G^{16}$.}

As an example, let us write down the form of the weight $-2$ Jacobi form induced by this procedure for an orbifold of 
$E_8 \times E_8$ heterotic string.
\begin{align}
 \phi_{-2,\lt}(\T, \z) = -\frac{1}{4 N \n^{18}} \sum_{a,b=1}^{N} &  k_{a b} \tht_1^{-2} \lp \T \frac{a}{N} 
	+ \frac{b}{N} \Big| \T \rp
      q^{(a/N)^2 ( \vec{\g}^2 -2) / 2} \notag \\
      &\prod_{k=1}^{s} y_k^{a \vec{\g} . \beta_k / N} \chi^{E8}_{1} \chi^{E8}_{2},
\end{align}
where $\chi^{E8}_{j}$ for $j=1,2$ is defined as
\begin{equation}
 \chi^{E8}_{j} = \sum_{i=1}^{4} \lp  \prod_{n=1+8(j-1)}^{8+8(j-1)} \tht_i \lb  
      \frac{(a \T + b) \vec{\g}[n] }{N}  + \sum_{r=1}^{s} z^r \vec{\beta_r}[n]         
	  \Bigg| \T  \rb  \rp,
\end{equation}
to accomplish the sum over $E8$ lattice points.

The easiest way to find $\psi_{0,\lt} (\T, \z)$ is to add $\vec{Q}$ as another basis vector to the basis of $\lt$. Then, after
computing the weight $-2$ Jacobi form
with this modified lattice as described above, one can find the effect of $Q^2$ insertion by differentiating with respect
to the variable $z^Q$ twice and then setting this variable to zero. After this, with an appropriate subtraction 
of $E_2 \phi_{-2,\lt}$, 
$\psi_{0,\lt} $ 
can be found. In other words, if we suppose $\hat{\phi}_{-2,\lt}(\T, \z, \z_Q)$ (so that
$\phi_{-2,\lt}(\T, \z) = \hat{\phi}_{-2,\lt}(\T, \z, \z_Q = 0)$)
is the weight $-2$ Jacobi form that also counts $\vec{Q}$ charges, $\psi_0 (\T, \z)$ is given by
\begin{equation}
 \psi_0 (\T, \z) = \frac{24}{2 \vec{Q}.\vec{Q}} \lb \lp y^Q \frac{\del}{\del y^Q} \rp^2 \hat{\phi}_{-2,\lt} \rb 
 \Bigg|_{ \z_Q = 0}
      - E_2 \phi_{-2,\lt}.
\end{equation}

\subsection{Symmetry Breaking in \texorpdfstring{$\mathcal{N}=2$}{N=2} Gauge Theories}
In this section, we will discuss how to break gauge symmetries in the $\mathcal{N}=2$ theories we consider here and thereby
move in the hypermultiplet moduli space. Let us suppose that we start with a $\mathcal{N}=2$ theory which has $\mf{g}$ as its gauge
Lie algebra where $\mf{g}$ is a direct sum of a semisimple part and a number of abelian Lie algebras. Remember that in our case
gauge symmetry is generated in 6-dimensions by compact scalars on a lattice $\lt$. 

There may be flat directions in the scalar potential of this theory that allows charged hypermultiplet scalars
to gain vev. 
Such a vev breaks gauge symmetry to a Lie subalgebra, $\mf{h}$, and reduces the rank by one.
We will
describe one such flat direction which we will use in following sections.
Details can be found in \cite{Danielsson:1996uv} and \cite{Honecker:2006qz}.

Each hypermultiplet consists of two chiral multiplets if we use the language of $\mathcal{N}=1$ supersymmetry.
These chiral multiplets have two scalars, say $\phi$ and $\phi^*$, which are CPT conjugates of each other.
In the flat direction we use for gauge symmetry breaking,
one needs two hypermultiplets with scalars $( \phi_1, \phi^*_1)$ and $( \phi_2, \phi^*_2)$ such that their $\phi$
components have charges $|\vec{q} \rangle$ and  $|-\vec{q} \rangle$ with respect to the Cartan subalgebra. 
Further, assuming that $|\vec{q} \rangle$ can not be set into the direction of $|-\vec{q} \rangle$ by a gauge transformation
in the non-abelian part, one can turn on vevs of the form $\phi_1, \phi^*_2  \sim v $. Out of these two hypermultiplets,
one linear combination gets mass together with the vector multiplet corresponding to the $U(1)$ generator in $\vec{q}$'s direction, and the 
other remains massless. 

Now, let $\vec{\alpha}$ be a root in the non-abelian part. 
A vector field $A^\mu_{\vec{\alpha}}$ corresponding to this root gets mass
through the minimal gauge coupling if the transformation matrix $t_\alpha$ acting on $|\pm \vec{q} \rangle$ 
(with respect to an appropriate representation of the non-abelian part) gives a non-vanishing result. Since, the supersymmetry
is unbroken, the whole vector multiplet gets mass together with this field. Furthermore, in this case,
the hypermultiplet $t_\alpha |\pm \vec{q} \rangle \sim |\pm \vec{q} + \alpha \rangle$ also gets mass through the quartic
scalar potential. Moreover, these are the only vector and hypermultiplets getting mass.
Such $|\pm \vec{q} \rangle$  pairs can be found if one has a hypermultiplet in a $C + \bar{C}$ representation (where $C$
is complex) or if there is a hypermultiplet in a real representation $R$ such that there is no root connecting
$| \vec{q} \rangle$  to $|- \vec{q} \rangle$ . 

There is a simple way to see whether a gauge symmetry $\mf{g}$ can be broken through
this procedure to a Lie subalgebra, $\mf{h}$, with its rank reduced by one. One starts with the adjoint of $\mf{g}$
and matter representations that will be used in the breaking (either $C + \bar{C}$ or $R$). Then, they are
decomposed with respect to the representations of a maximal $\mf{h} + \mf{u}(1)$ Lie subalgebra of $\mf{g}$. Firstly, there should
be singlets of $\mf{h}$ in the matter decomposition which are also charged under the $\mf{u}(1)$. We will denote these by
$\rep{1}_{\pm q}$. Moreover, for every factor, $\rep{R_i}_{q_i}$, in the decomposition of $\mf{g}$'s adjoint as
\begin{equation}
 \rep{Adj_g} \rightarrow \rep{1}_0 + \rep{Adj_h}_0 + \sum_i \rep{R_i}_{q_i},
\end{equation}
there should be factors in the matter decomposition in either of $\rep{R_i}_{\pm q + q_i}$ but not both. 
Then, vector multiplets in $\sum_i \rep{R_i}_{q_i}$ representations get mass together with corresponding hypermultiplets.
This gives
a sufficient condition for $\mf{g} \rightarrow \mf{h}$ breaking. Now, let us go over some examples which will be useful
in the coming sections as well.

\subsubsection*{$\mathbf{SU(n+1) \rightarrow SU(n)}$}
Start with $\mf{g} = \mf{su}_{n+1}$, where simple roots are given in an orthonormal basis as
\begin{equation}
 \vec{\beta}_1 = (1, -1, 0, \ldots, 0), \vec{\beta}_2 = (0, 1, -1, 0, \ldots, 0), \ldots,  \vec{\beta}_n = (0, \ldots, 0, 1, -1).
\end{equation}
We will display the familiar breaking pattern to $\mf{h} = \mf{su}_n$ by a $\rep{n+1} + \rep{\overline{n+1}}$ pair.
Take the simple roots of $\mf{su}_n$ to be $\vec{\beta}_1, \ldots, \vec{\beta}_{n-1}$. So, the sublattice of 
$\mf{su}_{n+1}$'s
root lattice which is orthogonal to $\mf{su}_n$'s root lattice is generated by $\vec{\beta} = (1, \ldots, 1, -n)$.\footnote{Though not 
important for this example, finding the surviving sublattice of $\lt$ as we break gauge symmetry will be important when
$U(1)$ factors are involved in the initial and/or final gauge group.} Then we see the following decompositions for 
$\mf{g} \rightarrow \mf{h} + \mf{u}(1)$ where $\mf{u}(1)$ is generated by $\vec{\beta}$:
\begin{equation}
 \rep{Adj^{n+1}} \rightarrow \rep{1}_0 + \rep{Adj^n}_0 + \rep{n}_{n+1} + \rep{\overline{n}}_{-n-1}, 
\end{equation}
and
\begin{equation}
 \rep{n+1} + \rep{\overline{n+1}} \rightarrow \lb \rep{1}_{-n} + \rep{n}_1 \rb 
						+  \lb \rep{1}_{n} + \rep{\overline{n}}_{-1} \rb.
\end{equation}
Now, giving vev to the scalars in $\rep{1}_{-n} + \rep{1}_{n}$ breaks the $U(1)$ symmetry. Moreover, gauge symmetry
is completely reduced to $SU(n)$ as hypermultiplets $\rep{n}_1 + \rep{\overline{n}}_{-1}$ gain mass together with vector
multiplets $ \rep{n}_{n+1} + \rep{\overline{n}}_{-n-1}  $. 

In summary, if one has a hypermultiplet pair 
$\rep{n} + \rep{\overline{n}}$, $SU(n)$ can be broken down to $SU(n-1)$. After breaking, what remains in the matter 
spectrum can be exemplified by
\begin{align}
 \rep{n} + \rep{\overline{n}} &\rightarrow \rep{1} \text{ if it is the symmetry breaking representation (rep.)}, \\
 \rep{n}  &\rightarrow \rep{1} + \rep{n-1} \text{ otherwise}, \\
 \rep{Asy^n} &\rightarrow \rep{n-1} + \rep{Asy^{n-1}}.
\end{align}
For reference, we also gave the decomposition of the antisymmetric representation, $\rep{Asy^n}$, as well.

Now, we give some more examples for later reference.

\subsubsection*{$\mathbf{SU(n) \rightarrow SU(n-2) \times SU(2)}$}
This can be accomplished by a $\rep{Asy^n} + \rep{\overline{Asy^n}}$ pair. Some examples of matter spectrum after breaking
are
\begin{align}
 \rep{Asy^n} + \rep{\overline{Asy^n}} &\rightarrow \rep{1,1} 
		    + \rep{Asy^{n-2},1} + \rep{\overline{Asy^{n-2}},1} \text{ if breaking rep.}, \\
 \rep{Asy^n}  &\rightarrow \rep{1,1} + \rep{Asy^{n-2},1} + \rep{n-2,2}, \\
 \rep{n} &\rightarrow \rep{n-2,1} + \rep{1,2}.
\end{align}

Note that, from the breaking representation a $\rep{Asy^{n-2},1} + \rep{\overline{Asy^{n-2}},1}$ pair survives.
So, if one starts with $SU(2N)$ such that there is a $\rep{Asy^{2N}} + \rep{\overline{Asy^{2N}}}$ hypermultiplet pair
in the matter spectrum, one can use this
to break the gauge symmetry down to $SU(2)^N$. Furthermore, if the initial spectrum contains two 
$\rep{Asy^{2N},1} + \rep{\overline{Asy^{2N}},1}$ pairs, the final spectrum contains matter representations
\begin{equation}
 2 \sum_{1 \leq n < k \leq N} \rep{2^n 2^k},
\end{equation}
where $\rep{2^n}$ is a doublet with respect to the $n^{th}$ $SU(2)$.

\subsubsection*{$\mathbf{SU(2)^N \rightarrow U(1)}$}
We take the lattice, $\lt$, to be created by simple roots
\begin{equation}
 \vec{\beta}_1 = (\sqrt{2}, 0, \ldots, 0), \ldots,  \vec{\beta}_N = (0, \ldots, 0, \sqrt{2}).
\end{equation}
We claim that it is possible to break $SU(2)^N$ down to a $U(1)$ so that the sublattice of $\lt$
in this $U(1)$'s direction is created by
\begin{equation}
\vec{\beta}_1 + \ldots +  \vec{\beta}_N = (\sqrt{2}, \ldots, \sqrt{2}).
\end{equation}
Note that this sublattice is $\langle 2N \rangle$, and hence if we can find such a breaking pattern in heterotic string 
theories we consider, the modified index will give rise to a weight $-2$, index $N$ Jacobi form.

The representation we will use for that symmetry breaking is $2 \sum_{1 \leq n < k \leq N} \rep{2^n 2^k}$ so that with
respect to the unbroken $U(1)$ we will be left with the following (this is for the symmetry breaking matter representation)
\begin{equation}
 2 \sum_{1 \leq n < k \leq N} \rep{2^n 2^k} \rightarrow (2N-1)(N-1) \rep{1}_0 + N(N-2) \rep{1}_2 + N(N-2) \rep{1}_{-2}.
\end{equation}
We will prove this assertion by induction.

For $N=2$, we start with the following representations in the vector and hyper multiplets:
\begin{equation}
 \textit{Vectors: } \rep{3^1} + \rep{3^2}, \text{ and } \textit{Hypers: } 2 \rep{2^1 2^2}.
\end{equation}
Decomposing the representations into the representations of $U(1)'s$ generated by $\vec{\beta}_1 + \vec{\beta}_2$ 
and the orthogonal $\vec{\beta}_1 - \vec{\beta}_2 $.
\begin{equation}
 \textit{Vectors: } 2 \rep{1}_{0,0} + \rep{1}_{2,2} + \rep{1}_{-2,-2} + \rep{1}_{2,-2} + \rep{1}_{-2,2},
\end{equation}
\begin{equation}
 \textit{Hypers: }  2 \rep{1}_{2,0} + 2 \rep{1}_{-2,0} + 2 \rep{1}_{0,2} + 2 \rep{1}_{0,-2}.
\end{equation}
Giving vev to the scalars of a $\rep{1}_{0,2} +  \rep{1}_{0,-2} $ pair, one can break down to the first $U(1)$. What remains
in the matter spectrum is three $\rep{1}_0$ representations, consistent with the claim.

Now, suppose that the claim is correct up to some $N$. We will show that it is also correct for $N+1$.
We start with
\begin{equation}
 \textit{Vectors: } \sum_{n=1}^{N+1} \rep{3^n}, \text{ and } \textit{Hypers: }  \sum_{1 \leq n < k \leq N+1} 2 \rep{2^n 2^k}.
\end{equation}
By the inductive hypothesis we can break this down to a $SU(2) \times U(1)$ where $U(1)$ is generated by,
say $ \vec{\beta} = \vec{\beta}_1 + \ldots +  \vec{\beta}_N$. Then, at this stage, the spectrum is
\begin{equation}
 \textit{Vectors: } \rep{3}_0 + \rep{1}_0,
\end{equation}
\begin{equation}
 \textit{Hypers: } (2N-1)(N-1) \rep{1}_0 + N(N-2) \rep{1}_2 + N(N-2) \rep{1}_{-2} + 2N \rep{2}_1 + 2N \rep{2}_{-1}.
\end{equation}
Under the $U(1) \times U(1)$ generated by $ \vec{\beta} + \vec{\beta}_{N+1}$ and orthogonal
$ \vec{\beta} - N \vec{\beta}_{N+1}$,  these representations decompose as
\begin{equation}
 \textit{Vectors: } 2 \rep{1}_{0,0} + \rep{1}_{2,-2N} + \rep{1}_{-2,2N} ,
\end{equation}
\begin{align}
 \textit{Hypers: } &(2N-1)(N-1) \rep{1}_{0,0} + N(N-2) \rep{1}_{2,2} + N(N-2) \rep{1}_{-2,-2} \notag\\
		 &+ 2N \lb \rep{1}_{2,-N+1} + \rep{1}_{0,N+1} + \rep{1}_{0,-N-1} + \rep{1}_{-2,N-1}  \rb.
\end{align}
Now, one can verify that giving vev to a $ \rep{1}_{0,N+1} + \rep{1}_{0,-N-1} $ 
pair breaks the gauge symmetry down to the first $U(1)$ generated by 
$\vec{\beta}_1 + \ldots +  \vec{\beta}_{N+1}$ with remaining matter representations
\begin{equation}
 \textit{Hypers: } (2N+1)N \rep{1}_{0} + (N-1)(N+1) \rep{1}_{2} + (N-1)(N+1) \rep{1}_{-2},
\end{equation}
consistent with the result we wanted to prove.

\subsection{An Orbifold Example}
Our main motivation for the last section was to find possible gauge symmetry
breaking patterns, supposing we start with a heterotic string model
at the orbifold limit. We can, then, make use of this orbifold limit to explicitly compute the Jacobi forms $\phi_{-2,\lt}$
and $\psi_{0,\lt} $ (and later to compute threshold corrections and also Gromov-Witten invariants for possible Type IIA duals).
The fact that the vector space of Jacobi forms is finite dimensional will sometimes enable us to write down the whole
Jacobi form by only using the massless spectrum of the six dimensional theory.

A rich example we will study is the $E_8 \times E_8$ heterotic string on a $T^4 / \ZZ_6$ orbifold for which the shift vector is
\begin{equation}
 \vec{\g} = (5, 1^7 ; 3^2, 0^6),
\end{equation}
where its components are given in an orthonormal basis. We choose the orthonormal basis so that the coordinates of a single $E_8$ lattice in this basis
are either all integral or all half-integral and are also constrained to have an even integer sum.

The gauge group surviving the orbifold projection is $SU(9) \times SU(2) \times E_7$.  A set of simple roots for these
gauge groups can be given as in the following, where we use the same basis in which $\vec{\g}$ components are given:
\begin{equation}
\begin{array}{ c ccc}
 SU(9): \text{  } & \vec{\alpha}_1 = (  0, 1, -1, 0^{13}  ),  
		   &\vec{\alpha}_2 = (  0^2,  1, -1, 0^{12}  ), 
		  & \vec{\alpha}_3 = (  0^3,  1, -1, 0^{11}  ),   \\
		   &\vec{\alpha}_4 = (  0^4,  1, -1, 0^{10}  ),   
		  & \vec{\alpha}_5 = (  0^5,  1, -1, 0^{9}  ),   
		   &\vec{\alpha}_6 = (  0^6,  1, -1, 0^{8}  ),   \\
		  & \vec{\alpha}_7 = (  1/2, -1/2^6, 1/2, 0^8  ),  
		  & \vec{\alpha}_8 = (  1/2^8,  0^8   ),   

\end{array}
\end{equation}
\begin{equation}
\begin{array}{ c ccc}
 SU(2): \text{  } & \vec{\alpha}_9 = (  0^8, 1, 1, 0^{6}  ),  & & \\
   E_7: \text{  }  &\vec{\alpha}_{10} = (  0^{8}, 1/2, -1/2^6, 1/2 ), 
		  & \vec{\alpha}_{11} = (   0^{14},  1, -1  ),   
		   &\vec{\alpha}_{12} = (  0^{13},  1, -1, 0^{1}  ),   \\
		  & \vec{\alpha}_{13} = (  0^{12},  1, -1, 0^{2}  ),   
		   &\vec{\alpha}_{14} = ( 0^{11},  1, -1, 0^{3}  ),   
		  & \vec{\alpha}_{15} = ( 0^{10},  1, -1, 0^{4} ),\\  
		  & \vec{\alpha}_{16} = (  0^{14},  1, 1  ).
\end{array}
\end{equation}
 
We further compactify  this theory on a $T^2$ to get a $\mc{N} =2$ theory in four dimensions. 
The massless spectrum for this particular example,
when all Wilson lines are switched off, is given by (with respect to the $SU(9) \times SU(2) \times E_7$ gauge group)
\begin{equation}
 \textit{Vectors: }  \rep{80,1,1} + \rep{1,3,1} + \rep{1,1,133},
\end{equation}
\begin{align}
 \textit{Hypers: } &2 \rep{1,1,1} + \rep{9,2,1} + \rep{\bar{9},2,1} + 2 \rep{36,1,1} + 2\rep{\bar{36},1,1} \notag\\
		   &+5 \rep{9,1,1} + 5 \rep{\bar{9},1,1} + 3\rep{1,1,56} + 10 \rep{1,2,1} ,
\end{align}
where we have written the matter representations in a manifestly real fashion. Among the matter multiplets,
 $2 \rep{9,2,1}$ comes from $\ZZ_6$ fixed points, $4 \rep{36,1,1} + 10 \rep{9,1,1}$ comes from $\ZZ_3$ fixed points and
 $3\rep{1,1,56} + 10 \rep{1,2,1} $ comes from $\ZZ_2$ fixed points.

 The massless spectrum has some interesting features which makes it a useful example for our purposes. 
 Firstly, $E_7$ can be broken independently of $SU(9) \times SU(2)$ by giving a vev to 
 hypermultiplet scalars in the $3\rep{1,1,56}$ and through the following
 chain:
 \begin{equation}
\begin{array}{ccc}
 G: E_7 &  \textit{Vectors: } \rep{133}  &  \textit{Hypers: } 3\rep{56}, \\
 G: E_6 &  \textit{Vectors: } \rep{78}  &  \textit{Hypers: } 5\rep{1} + 2 \rep{27} + 2 \rep{\overline{27}}, \\
 G: SO(10) &  \textit{Vectors: } \rep{45}  &  \textit{Hypers: } 8\rep{1} + 4\rep{10} +  \rep{16} + \rep{\overline{16}}, \\
 G: SU(5) &  \textit{Vectors: } \rep{24}  &  \textit{Hypers: } 9\rep{1} + 5 \rep{5} + 5 \rep{\overline{5}}, \\
 G: SU(4) &  \textit{Vectors: } \rep{15}  &  \textit{Hypers: } 18 \rep{1} + 4 \rep{4} + 4 \rep{\overline{4}}, \\
 G: SU(3) &  \textit{Vectors: } \rep{8}  &  \textit{Hypers: } 25 \rep{1} + 3 \rep{3} + 3 \rep{\overline{3}}, \\
 G: SU(2) &  \textit{Vectors: } \rep{3}  &  \textit{Hypers: } 30 \rep{1} + 4 \rep{2},  \\
 G: 1 &  \textit{Vectors: } -  &  \textit{Hypers: } 35 \rep{1}.  \\
\end{array}
\end{equation}
Note that this is the same spectrum as one would have, if $E_7$ symmetry is broken in a smooth compactification with $10$
instantons on one side of the $E_8$'s \cite{Aldazabal:1995yw}. In our examples below  we will usually assume that this $E_7$
is completely broken and hence no Wilson lines will be switched on for groups in this chain. Only if we work out the 
weight zero Jacobi form, $\psi_{0,\lt}$, we will assume a $SU(2)$ is left unbroken from the $E_7$ chain.
So, for definiteness, we fix $\vec{\beta}_Q$ to be $\vec{\alpha}_{11}$ whenever we compute $\psi_{0,\lt}$ for this model.
Note that,
even in this case, we will assume that the Wilson line (or more appropriately 
the complex modulus, $V^{Q}$) for this $SU(2)$ is zero.
 
Now, if $E_7$ is completely broken by the chain described above, one gets the following spectrum under $SU(9) \times SU(2)$:
\begin{equation}
 \textit{Vectors: }  \rep{80,1} + \rep{1,3},
\end{equation}
\begin{equation}
 \textit{Hypers: } 37 \rep{1,1} + \rep{9,2} + \rep{\bar{9},2} + 2 \rep{36,1} + 2\rep{\bar{36},1} 
		   +5 \rep{9,1} + 5 \rep{\bar{9},1} + 10 \rep{1,2} .
\end{equation}
Looking at the massless spectrum, we see that $SU(2)$ can be broken by using the 
two hypermultiplets in the $\rep{1,2}$ representation.
$SU(9)$ can also be broken completely by using the $\rep{9,1} + \rep{\bar{9},1}$ hypermultiplets
and through chains of $SU(N) \rightarrow SU(N-1)$. Note that antisymmetric
representations $2 \rep{36,1} + 2\rep{\bar{36},1}$ provide two new $N + \bar{N}$ pairs
at each step of $SU(N) \rightarrow SU(N-1)$,
compensating the loss of a single pair due to symmetry breaking.

Furthermore, $\rep{36,1} + \rep{\bar{36},1}$ can be used to break $SU(9)$ to $SU(N) \times \lb SU(2) \rb^n$ form. Also,
since there are two such pairs, one obtains $2 \rep{2^i 2^j}$ representations as described in the previous section.
These $2 \rep{2^i 2^j}$ representations then can be used to break $\lb SU(2) \rb^n$ to a $U(1)$ with 
$\lt = \langle 2n \rangle$. Interestingly, if the original $SU(2)$ of the spectrum is kept unbroken, it also provides 
hypermultiplets of the form $2 \rep{2^i 2^j}$ via $\rep{9,2} + \rep{\bar{9},2}$ hypermultiplets.

Now, one can apply these three symmetry breaking patterns in varying orders to get a large class of theories with 
gauge symmetries of the form $SU(N) \times \lb SU(2) \rb^n \times \lb U(1) \rb^k$. We now give several such examples.

\subsubsection*{Example 1 : $\mathbf{\lt = A_1}$}
The gauge symmetry can be broken down to a $SU(2)$, where the $SU(2)$ comes either from the original $SU(2)$ or from
the $SU(9)$ factor. In both cases, the massless spectrum is
\begin{equation}
 \textit{Vectors: }  \rep{3},
\end{equation}
\begin{equation}
 \textit{Hypers: } 191 \rep{1} + 28 \rep{2}.
\end{equation}
Now, turning on Wilson lines and going to a generic point on the vector multiplet moduli space gives a theory with
$(N_v, N_h - 1) = (4, 190)$, where $N_v$ is the number of massless vector multiplets and $N_h$ is the number of massless
hypermultiplets in four dimensions.

The spectrum above fixes the first two Fourier coefficients of $\phi_{-2}(\T, \z)$ as
\begin{equation}
 \phi_{-2,1} = -\frac{2}{q} + 2 \lp - y_1^{-2} + 28 y_1^{-1} + 186 + 28 y_1 - y_1^2  \rp  + O(q).
\end{equation}
The coefficient of $q^0 y_1^0$ term, $186$, is fixed by noting that the gravity multiplet contributes with a $-1$ to this term
and the contributions from hypermultiplets and vector multiplets are $191$ and $-4$, respectively. 
This fixes the full Jacobi form to be 
\begin{equation}\label{eq:m1}
 \phi_{-2,1} = - \frac{2}{\Delta(q)} \frac{14 E_4(q) E_{6,1}(q, y_1) + 10 E_{4,1}(q, y_1) E_6(q)}{24},
\end{equation}
where the Jacobi-Eisenstein series $E_{6,1}$ and $E_{4,1}$ are defined in the Appendix. An explicit computation at the orbifold
point verifies this.
Indeed, this example is very
widely studied
in the literature. It first appeared in \cite{LopesCardoso:1996nc}, where the fact that Jacobi forms arise in the computation
of threshold corrections with Wilson lines was also noted. Threshold corrections and its relation to Jacobi
forms for $\lt = A_1$ also appeared in \cite{Kawai:1996te}.

The numbers $14$ and $10$ are interpreted as the instanton numbers in a geometric compactification so that the remaining $SU(2)$
is in the $E_8$ factor with $14$ instantons. If the $SU(2)$ comes from the $SU(9)$ factor, this interpretation 
is also consistent with our case
since this orbifold model can be matched, after Higgsing, to a geometric compactification with $(14,10)$ instantons 
\cite{Stieberger:1998yi}. Of course, this argument can not be used if the unbroken $SU(2)$ is the original $SU(2)$ 
in the $SU(9) \times SU(2)$ model, since this $SU(2)$ is special to the orbifold limit and is broken if the orbifold is
blown up to a smooth compactification. Still, interestingly, this $SU(2)$ behaves as if it is an $SU(2)$ on the $14$ instanton
side of a geometric compactification of $E_8 \times E_8$ heterotic string, as far as threshold corrections are concerned.

Group theoretically, we can explain this symmetry in the massless spectrum between $SU(9)$ and $SU(2)$ representations
by considering a hypothetical 
theory with $SU(11)$ gauge symmetry and massless spectrum:
\begin{equation}
 \textit{Vectors: }  \rep{120},
\end{equation}
\begin{equation}
 \textit{Hypers: } 34 \rep{1} + 5 \rep{11} + 5 \rep{\overline{11}} + 2 \rep{55} + 2\rep{\overline{55}}.
\end{equation}
Breaking $SU(11)$ to $SU(9) \times SU(2)$ using $\rep{55} + \rep{\overline{55}}$ gives a massless spectrum precisely as in our
example coming from the orbifold limit.

\subsubsection*{Example 2 : $\mathbf{\lt = \langle 4 \rangle}$}
Now  we consider an example in which the final gauge group is a $U(1)$ obtained by breaking  $SU(2) \times SU(2)$ as
described in the previous section. The massless hypermultiplet spectrum is given by
\begin{equation}
 \textit{Hypers: } 149 \rep{1}_0 + 48 \rep{1}_{1} + 48 \rep{1}_{-1}.
\end{equation}
When Wilson lines are turned on, this gives a theory with $(N_v, N_h -1) = (4, 148)$.
This gives the beginning of $\phi_{-2,2}$'s Fourier series as
\begin{equation}
 \phi_{-2,2} = -\frac{2}{q}+2 \left(48 y_1+\frac{48}{y_1}+144\right) + O(q).
\end{equation}
If we form the most general weight $-2$, index $2$ nearly holomorphic Jacobi form which starts as $-2/q$ then by
matching the Fourier coefficients for $q^{-1}$ and $q^0$ terms we can uniquely fix $\phi_{-2,2}$ as
\begin{equation}\label{eq:m2}
 \phi_{-2,2} = - \frac{2}{\Delta(q)} \frac{1}{6 \times 12^2} \Big( 6  \tilde{\phi}_{0,1}^2 E_6 E_4 
    -7 \tilde{\phi}_{-2,1} \tilde{\phi}_{0,1} E_4^3 
    -  5 \tilde{\phi}_{-2,1} \tilde{\phi}_{0,1} E_6^2 
    + 6  \tilde{\phi}_{-2,1}^2 E_6 E_4^2 
 \Big).
 \end{equation}
$\tilde{\phi}_{-2,1}$ and $ \tilde{\phi}_{0,1}$ are the generators of  
even weight weak Jacobi forms over lattices $\lt = \langle 2m \rangle$, with $m \in \ZZ^+$, 
when considered as a ring over modular forms \cite{Zagier}. Detailed definitions can be found in the Appendix.

We have checked this to order $O(q^7)$ by comparing it to the orbifold computation where we take 
$\lt = \langle \vec{\alpha}_1 + \vec{\alpha}_3 \rangle$. Up to order $q$, the result is given by
\begin{align}
 \phi_{-2,2} = &-\frac{2}{q}+\left(96 y_1+\frac{96}{y_1}+288\right)+q \Big(-2 y_1^4+96 y_1^3+10192 y_1^2+69280
   y_1  \notag\\
   &+\frac{69280}{y_1}+\frac{10192}{y_1^2}+\frac{96}{y_1^3}-\frac{2}{y_1^4}+123756 \Big)
   +O\left(q^{2}\right).
\end{align}

Threshold corrections for a $(N_v, N_h -1) = (4, 148)$ model (though in a smooth compactification
with $(13,11)$ instanton embeddings) and its relation
to weight $-2$, index $2$ Jacobi forms appeared in \cite{LopesCardoso:1996zj}. An indirect argument
is used there to get 
information about the Jacobi form $ \phi_{-2,2} $ similar to the argument we have given above. However, instead of using Jacobi forms,
a set of Siegel forms with correct singularity structures are matched to the expressions arising in threshold corrections.
We will describe the details of threshold corrections in the next section.

\subsubsection*{Example 3 : $\mathbf{\lt = \langle 6 \rangle}$}
In this example, the final gauge group is a $U(1)$ obtained by breaking a $\lb SU(2) \rb^3$. 
The massless hypermultiplet spectrum for zero Wilson lines is given by
\begin{equation}
 \textit{Hypers: } 119 \rep{1}_0 + 60 \rep{1}_{1} + 60 \rep{1}_{-1} + 3 \rep{1}_2 + 3 \rep{1}_{-2}.
\end{equation}
When Wilson lines are turned on, we get $(N_v, N_h -1) = (4, 118)$ massless multiplets.
The beginning of $\phi_{-2,3}$'s Fourier series is given as
\begin{equation}
 \phi_{-2,3} = -\frac{2}{q}+2 \left(3 y_1^2+\frac{3}{y_1^2} + 60 y_1+\frac{60}{y_1}+114\right) + O(q).
\end{equation}
Again by forming the most general nearly holomorphic Jacobi form which starts with
the same Fourier coefficients for $q^{-1}$ and $q^0$ terms and has weight $-2$, index $3$,
we can uniquely fix $\phi_{-2,3}$ as
\begin{align}\label{eq:m3}
 \phi_{-2,3} = &- \frac{2}{\Delta(q)} \frac{1}{4 \times 12^3} \Big( 4  \tilde{\phi}_{0,1}^3 E_6 E_4 
    -5 \tilde{\phi}_{-2,1} \tilde{\phi}_{0,1}^2 E_6^2 
    + 12 \tilde{\phi}_{-2,1}^2 \tilde{\phi}_{0,1} E_6 E_4^2 \notag\\
    &-3 \tilde{\phi}_{-2,1}^3 E_6^2 E_4
    -7 \tilde{\phi}_{-2,1} \tilde{\phi}_{0,1}^2 E_4^3
    - \tilde{\phi}_{-2,1}^3  E_4^4
 \Big).
 \end{align}

This can also be compared with the explicit orbifold computation
 for which we matched the coefficients to order $O(q^7)$. For the orbifold computation, we have taken 
$\lt = \langle \vec{\alpha}_1 + \vec{\alpha}_3 + \vec{\alpha}_5 \rangle$. Up to order $q$, the result is
\begin{align}
 \phi_{-2,3} = &-\frac{2}{q}+\left(6 y_1^2+120 y_1+\frac{120}{y_1}+\frac{6}{y_1^2}+228\right)+q 
 \Big(6y_1^4+1776 y_1^3+20292 y_1^2 \notag\\
 &+69072 y_1 
 +\frac{69072}{y_1}+\frac{20292}{y_1^2}+\frac{1776}
 {y_1^3}+\frac{6}{y_1^4}+100596\Big)
   +O\left(q^{2}\right).
\end{align}

At this point we should note that these three examples can be obtained from an intermediate $SU(6)$ model in the breaking 
chain, which has the same massless spectrum as in \eqref{eq:SU6model}. This model can be obtained as a smooth 
six dimensional heterotic
compactification as described in \cite{Bershadsky:1996nh} where a F-theory dual is also described.

\subsubsection*{Example 4 : $\mathbf{\lt = \langle 8 \rangle}$}
In this example, the final gauge group is a $U(1)$ obtained from a $\lb SU(2) \rb^4$ step. 
The massless hypermultiplet spectrum of the six dimensional theory is given by
\begin{equation}
 \textit{Hypers: } 101 \rep{1}_0 + 64 \rep{1}_{1} + 64 \rep{1}_{-1} + 8 \rep{1}_2 + 8 \rep{1}_{-2}.
\end{equation}
When Wilson lines are turned on, we get $(N_v, N_h -1) = (4, 100)$ massless multiplets.
The Jacobi form $\phi_{-2,4}$'s Fourier series start as
\begin{equation}
 \phi_{-2,4} = -\frac{2}{q}+2 \left(8 y_1^2+\frac{8}{y_1^2} + 64 y_1+\frac{64}{y_1}+96\right) + O(q).
\end{equation}
Again, we form the most general nearly holomorphic Jacobi form with appropriate weight and index.
Then, we uniquely fix $\phi_{-2,4}$ as
\begin{align}\label{eq:m4}
 \phi_{-2,4} = &- \frac{2}{\Delta(q)} \frac{1}{3 \times 12^4} \Big( 3  \tilde{\phi}_{0,1}^4 E_6 E_4
 -5 \tilde{\phi}_{-2,1} \tilde{\phi}_{0,1}^3 E_6^2
 +18 \tilde{\phi}_{-2,1}^2 \tilde{\phi}_{0,1}^2 E_6 E_4^2
 -9 \tilde{\phi}_{-2,1}^3 \tilde{\phi}_{0,1} E_6^2 E_4 \notag \\
 &+2 \tilde{\phi}_{-2,1}^4   E_6^3
 -7 \tilde{\phi}_{-2,1} \tilde{\phi}_{0,1}^3 E_4^3
 -3 \tilde{\phi}_{-2,1}^3 \tilde{\phi}_{0,1} E_4^4
 + \tilde{\phi}_{-2,1}^4 E_6 E_4^3
 \Big).
 \end{align}

We also perform the computation at the orbifold limit by taking 
$\lt = \langle \vec{\alpha}_1 + \vec{\alpha}_3 + \vec{\alpha}_5 + \vec{\alpha}_7 \rangle$. 
Up to order $q$, the result is
\begin{align}
 \phi_{-2,4} = &-\frac{2}{q}+\left(16 y_1^2+128 y_1+\frac{128}{y_1}+\frac{16}{y_1^2}+192\right)+
 q \Big(228 y_1^4+4992 y_1^3+26880 y_1^2 \notag\\
   &+ 65664 y_1+\frac{65664}{y_1}+\frac{26880}{y_1^2}+\frac{4992}{y_1^3}+\frac{228}{y_1^4}+87360\Big)
   +O\left(q^{2}\right).
\end{align}

\subsubsection*{Example 5 : $\mathbf{\lt = \langle 10 \rangle}$}
As noted before, we can also break $SU(9) \times SU(2)$ to a  $\lb SU(2) \rb^5$ first
and then use matter representations $ 2 \rep{2^i 2^j}$ to get a $U(1)$ with $\lt = \langle 10 \rangle$ . 
The massless hypermultiplet spectrum of the six dimensional theory is given by
\begin{equation}
 \textit{Hypers: } 95 \rep{1}_0 + 60 \rep{1}_{1} + 60 \rep{1}_{-1} + 15 \rep{1}_2 + 15 \rep{1}_{-2}.
\end{equation}
In the four dimensional theory, when Wilson lines are turned on, we get $(N_v, N_h -1) = (4, 94)$ massless multiplets.
The Jacobi form $\phi_{-2,5}$'s Fourier series for this example begin by
\begin{equation}
 \phi_{-2,5} = -\frac{2}{q}+2 \left(8 y_1^2+\frac{8}{y_1^2} + 64 y_1+\frac{64}{y_1}+90\right) + O(q).
\end{equation}
From the most general nearly holomorphic Jacobi form with appropriate weight and index we 
obtain $\phi_{-2,5}$ by looking only at the $q^{-1}$ and $q^0$ terms as
\begin{align}\label{eq:m5}
 \phi_{-2,5} = &- \frac{2}{\Delta(q)} \frac{1}{12^6} \Big( 12  \tilde{\phi}_{0,1}^5 E_6 E_4
 -35  \tilde{\phi}_{-2,1} \tilde{\phi}_{0,1}^4 E_4^3
 -25  \tilde{\phi}_{-2,1} \tilde{\phi}_{0,1}^4 E_6^2
 +120  \tilde{\phi}_{-2,1}^2 \tilde{\phi}_{0,1}^3 E_6 E_4^2 \notag\\
 &-30  \tilde{\phi}_{-2,1}^3 \tilde{\phi}_{0,1}^2  E_4^4
 -90  \tilde{\phi}_{-2,1}^3 \tilde{\phi}_{0,1}^2 E_6^2 E_4 
 + 20  \tilde{\phi}_{-2,1}^4 \tilde{\phi}_{0,1} E_6 E_4^3 \notag\\
  &+ 40 \tilde{\phi}_{-2,1}^4 \tilde{\phi}_{0,1} E_6^3 
   + 9  \tilde{\phi}_{-2,1}^5 E_4^5  
  -21  \tilde{\phi}_{-2,1}^5 E_6^2 E_4^2
 \Big).
 \end{align}

We also perform the computation at the orbifold point by taking 
$\lt = \langle \vec{\alpha}_1 + \vec{\alpha}_3 + \vec{\alpha}_5 + \vec{\alpha}_7 + \vec{\alpha}_9 \rangle$
and match the two results up to order $O(q^7) $ .
Up to order $q$, the result is
\begin{align}
 \phi_{-2,5} = &-\frac{2}{q}+\left(30 y_1^2+120y_1+\frac{120}{y_1}+\frac{30}{y_1^2}+180\right)+q 
 \Big(8 y_1^5+980 y_1^4+8520 y_1^3+30580 y_1^2 \notag\\
  &+62320y_1+
 \frac{62320}{y_1}+\frac{30580}{y_1^2}+\frac{8520}{y_1^3} +\frac{980}{y_1^4}+\frac{8}{y_1^5}+78072\Big)
   +O\left(q^{2}\right).
\end{align}

\subsubsection*{Examples 6 and 7 : $\mathbf{\lt = A_2 }$ and $\mathbf{\lt =  A_3 }$}
One can also break $SU(9) \times SU(2)$ to a $SU(3)$ or $SU(4)$. For the $SU(3)$ case, the massless hypermultiplets of 
the six dimensional theory are
\begin{equation}
 \textit{Hypers: } 162 \rep{1} + 15 \rep{3} + 15 \rep{\bar{3}},
\end{equation}
and for the $SU(4)$ case they are
\begin{equation}
 \textit{Hypers: } 139 \rep{1} + 12 \rep{4} + 12 \rep{\bar{4}} + 4 \rep{6}.
\end{equation}

This fixes the beginning of $\phi_{-2,A_2}$ and $\phi_{-2,A_3}$ as
\begin{align}
 \phi_{-2,A_2} = &-\frac{2}{q}+ \Big(-\frac{2 y_1^2}{y_2}-2 y_2 y_1+\frac{30y_1}{y_2}-\frac{2 y_1}{y_2^2}+30 y_1+
  30y_2+\frac{30}{y_2}-\frac{2 y_2^2}{y_1}+\frac{30y_2}{y_1} \notag\\
  &-\frac{2}{y_2 y_1}+\frac{30}{y_1}-
  \frac{2 y_2}{y_1^2}+312\Big),
\end{align}
and
\begin{align}
 \phi_{-2,A_3} = &-\frac{2}{q}+\Big(-\frac{2 y_1^2}{y_2}+\frac{8 y_3 y_1}{y_2}-\frac{2 y_3 y_1}{y_2^2}-2 y_3 y_1
 +\frac{24 y_1}{y_2}-\frac{2 y_2 y_1}{y_3}-\frac{2 y_1}{y_2 y_3}+\frac{8 y_1}{y_3}+24 y_1 \notag \\
 &-\frac{2 y_3^2}{y_2}
 +8 y_2+\frac{24 y_3}{y_2}+24 y_3+\frac{8}{y_2}+\frac{24 y_2}{y_3}+\frac{24}{y_3}
 -\frac{2 y_2}{y_3^2}+\frac{24 y_2}{y_1}-\frac{2 y_2 y_3}{y_1}-\frac{2 y_3}{y_2 y_1} \notag\\
 &+\frac{8 y_3}{y_1}-\frac{2 y_2^2}{y_3 y_1} 
 +\frac{8 y_2}{y_3 y_1}-\frac{2}{y_3 y_1}+\frac{24}{y_1}
 -\frac{2 y_2}{y_1^2}+264\Big)
   +O\left(q \right).
\end{align}
Note that we pick $\lt$ basis to be a set of simple roots and accordingly the exponents of $y_i$ are given by fundamental
weights associated with the irreducible representations in the spectrum.

Using the generators of Weyl invariant Jacobi forms for $A_2$ and $A_3$ root lattices (given in Appendix) and these two
expressions we can fix the full Jacobi form to be
\begin{equation}
 \phi_{-2,A_2} = - \frac{2}{\Delta(q)} \frac{1}{144} \Big( 6 \tilde{\phi}_{0,A_2} E_4 E_6
		   + 5 \tilde{\phi}_{-2,A_2} E_6^2 + 7 \tilde{\phi}_{-2,A_2} E_4^3 \Big),
\end{equation}
and
\begin{equation}
 \phi_{-2,A_3} = - \frac{2}{\Delta(q)} \frac{1}{864} \Big( 4 \tilde{\phi}_{0,A_3} E_4 E_6
		   + 5 \tilde{\phi}_{-2,A_3} E_6^2 + 7 \tilde{\phi}_{-2,A_3} E_4^3
		   - 8 \tilde{\phi}_{-4,A_3} E_4^2 E_6 \Big).
\end{equation}

We can also compute the Jacobi form $\phi_{-2,\lt}$ for these two cases using the orbifold limit. This orbifold computation
was essentially done in \cite{Weiss:2007tk}.
By taking 
$\lt = \langle \vec{\alpha}_1 , \vec{\alpha}_2 \rangle$ and 
$\lt = \langle \vec{\alpha}_1 , \vec{\alpha}_2, \vec{\alpha}_3 \rangle$ in our example we match the expressions
given above up to order $q^2$.

\subsection{Threshold Corrections}
In this section, we will discuss the perturbative gauge and gravitational coupling constants in 
the low energy $\mc{N} = 2$ effective field theory. 
Their dependence on momentum scale and vector multiplet moduli are given by (if the gauge 
symmetry is created at level 1) \cite{Antoniadis:1992sa, Antoniadis:1992rq, Kaplunovsky:1995jw}
\begin{equation}\label{eq:string_gauge}
 \frac{1}{g_\text{gauge}^2 (p^2)} = \Re \lp -\ii S + \frac{1}{\fpi} \Delta^\text{univ} \rp + 
     \frac{b_\text{gauge}}{\fpi} \log \frac{M_\text{str}^2}{p^2} + \frac{1}{\fpi} \Delta_\text{gauge},
\end{equation}
and
\begin{equation}\label{eq:string_grav}
 \frac{1}{g_\text{grav}^2 (p^2)} = 24 \Re \lp -\ii S + \frac{1}{\fpi} \Delta^\text{univ} \rp + 
     \frac{b_\text{grav}}{\fpi} \log \frac{M_\text{str}^2}{p^2} + \frac{1}{\fpi} \Delta_\text{grav}.
\end{equation}
Here, the one loop beta function coefficients
are given as
\begin{equation}
 b_\text{gauge} = 2 \Tr_\text{hyper}(Q^2) - 2 \Tr_\text{vector}(Q^2),
\end{equation}
\begin{equation}
  b_\text{grav} = 46 + 2 (N_h - N_v  ),
\end{equation}
and  $ \Delta^\text{univ}$ is coming from to the Green-Schwarz term.
We will mostly use the conventions of \cite{Henningson:1996jz} in this section.

In the previous sections, we described two Jacobi forms
associated with $ \Delta_\text{grav} $ and $ \Delta_\text{gauge} $: 
\begin{equation}
 \phi_{-2}(\T, \z) =  \sum_{n, k_i} c(n,k_i) q^n y_1^{k_1} \ldots y_s^{k_s} 
 = \sum_{\mu \in \lt^* / \lt} \tht_{\lt, \mu}(\T, \vec{z}) h_\mu(\T),
\end{equation}
and
\begin{equation}
 \psi_{0}(\T, \z) =  \sum_{n, k_i} d(n,k_i) q^n y_1^{k_1} \ldots y_s^{k_s} 
 = \sum_{\mu \in \lt^* / \lt} \tht_{\lt, \mu}(\T, \vec{z}) f_\mu(\T),
\end{equation}
from which we can compute gauge and gravitational threshold corrections by performing
the integrals in \eqref{eq:gravthre} and \eqref{eq:gaugethre}.

At this point, it will be useful to define a positivity notion on the lattice $\lt^*$ and more generally
on $\lt^* \otimes \RR$.
First we divide the space $\lt \otimes \RR = \lt^* \otimes \RR$ into two half spaces,
one positive and one negative. For our examples, we will use
a lexicographic ordering as follows. We decompose a vector, $\vec{v} \in \lt^* \otimes \RR$, to its components
with respect to the basis $\{ \vec{\gamma^i} \}$ as $\vec{v} = b_i \vec{\gamma^i}$. Then, we declare a nonzero vector, 
$\vec{v}$, positive if 
\begin{equation}
 b_1 = \ldots = b_{i-1} = 0 \text{ and } b_i > 0
\end{equation}
is satisfied for at least one of $i = 1, 2, \ldots, s$. Similarly, when we say $(b_1, \ldots, b_s) \in \ZZ^s$ is positive
we will mean that the same condition above is satisfied.

Now, following \cite{Kawai:1998md}, we define
\begin{equation}
 \mc{C}_0 = \frac{1}{2} \sum_{\vec{b}} c(0,\vec{b}),
\end{equation}
and
\begin{equation}
 \mc{C}_{2n} = \sum_{\vec{b} > 0}  c(0,\vec{b}) \lp b_i d^{i j} b_j \rp^n \text{ for } n \in \ZZ^+,
\end{equation}
where $c(.,.)$ are the Fourier coefficients of $\phi_{-2, \lt}$. We will also use 
$\tilde{\phi}_{0, \lt} = \phi_{-2, \lt} E_2$, and use a notation
$\tilde{\mc{C}}_{2n}$ defined in a similar way as above, but this time using the 
Fourier coefficients, $\tilde{c}(n,\vec{b})$, of $\tilde{\phi}_{0, \lt}$.
In the same way, we will write $\mc{D}_{2n}$ for similar sums with respect to the 
Fourier coefficients, $d(n,\vec{b})$, of $\psi_{0, \lt}$.

Lastly, we will talk about the elements $r_a = ( k, l, \vec{v} = b_i \vec{\gamma^i}) $
of $U \oplus \lt^*$ where $k, l, b_i \in \ZZ$. We will say $r$ is positive and write $r>0$ when
\begin{equation}
 k>0,
\end{equation}
\begin{equation}
\text{or } k = 0 \text{ and } l >0,
\end{equation}
\begin{equation}
\text{or } k = l = 0 \text{ and } \vec{v} > 0 \text{ (or equivalently } \vec{b} > 0 \text{)}.
\end{equation}
We define the vector with upper index as $r^a = \eta^{a b} r_b$ where $\eta^{a b}$ is the inverse of the metric
$\eta_{a b}$ satisfying $y^a \eta_{a b} y^b = -2 T U + V^i d_{ i j} V^j$. In the following,
we will mean $c(kl, \vec{b})$ when we write $c(r)$ and we will use the notation 
$r.y \equiv r_a y^a = k T + l U + b_i V^i $.

Integrals giving the threshold corrections,
\begin{equation}
\Delta_\text{grav} =  \int_\mc{F} \frac{\diff^2 \T}{\T_2} \lb 
    \lp E_2(q) - \frac{3}{\pi \T_2}  \rp  \sum_{\mu \in \lt^* / \lt} Z_{\G,\mu}(\T, T, U, V^i) h_\mu 
    - \tilde{c}(0) \rb, 
\end{equation}
and
\begin{equation}
24 \Delta_\text{gauge} - \Delta_\text{grav} =  \int_\mc{F} \frac{\diff^2 \T}{\T_2} \lb 
      \sum_{\mu \in \lt^* / \lt} Z_{\G,\mu}(\T, T, U, V^i) f_\mu 
    - d(0) \rb, 
\end{equation}
can be evaluated using the work of \cite{Borcherds:1996automorphic} which generalizes threshold correction integrals
in \cite{Harvey:1995fq} to a wide class of automorphic integrands. The result we need can be basically read from
\cite{Kawai:1998md}.\footnote{Rotational symmetry property of the Jacobi form as in equations (2.14) and
(2.15) of \cite{Kawai:1998md} is needed for the result given above to be correct and that can be easily checked
for the particular examples we work with. More general arguments can be given along the lines of 
\cite{Gritsenko:2012qn}.}

The gravitational threshold correction is given as
\begin{align}\label{eq:int_grav}
\Delta_\text{grav} = 4 \Re &\lp  \sum_{r>0} \lb \tilde{c}(r) \Li_1 \lp e^{2 \pi \ii r.y} \rp
			+ \frac{6}{\pi (y_2,y_2)} c(r) \mc{P}(r.y) \rb  \rp 
			+ \tilde{c}(0) \lb - \log\lp-(y_2,y_2)\rp - \mc{K} \rb \notag\\
      &+ \frac{6}{\pi^2 (y_2, y_2)} c(0) \zeta(3) + 4 \pi \rho_a y_2^a 
      +\frac{192 \pi}{(y_2, y_2)}\frac{1}{6} d_{a b c} y_2^a y_2^b y_2^c.
\end{align}
Here, $\mc{K} = \log \lb \frac{4 \pi}{\sqrt{27}} e^{1-\gamma_E}   \rb$, where $\gamma_E \approx 0.57721\ldots $ 
is the Euler - Mascheroni constant,
\begin{equation}
 \frac{1}{6} d_{a b c} y^a y^b y^c = -\frac{\mc{C}_4}{4s(s+2)} T V^i d_{ij} V^j + \frac{\mc{C}_0}{720} U^3
	      - \frac{\mc{C}_2}{24 s} U V^i d_{i j} V^j + \frac{1}{12} \sum_{\vec{b} > 0}  c(0,\vec{b}) \lp b_i V^i \rp^3,
\end{equation}
and
\begin{equation}
 \rho_a y^a = \lp \frac{2 \tilde{\mc{C}}_2}{s} - \frac{12 \mc{C}_4}{s(s+2)} \rp T + \frac{\tilde{\mc{C}}_0}{6} U
		-  \sum_{\vec{b} > 0}  c(0,\vec{b}) b_i V^i .
\end{equation}
Definitions for $\mc{P}(x)$ and polylogarithms, $\Li_n(x)$, are given in the Appendix.

To find the gauge threshold correction, we compute the difference
\begin{equation}\label{eq:int_gauge}
 24 \Delta_\text{gauge} - \Delta_\text{grav} = 8 \pi \kappa_a y_2^a + 
	    d(0) \lb - \log\lp-(y_2,y_2)\rp - \mc{K} \rb 
	    + 4   \sum_{r>0} d(r) \Re \lb \Li_1 \lp e^{2 \pi \ii r.y} \rp \rb,
\end{equation}
where
\begin{equation}
  \kappa_a y^a = \frac{\mc{D}_2}{s} T + \frac{\mc{D}_0}{12} U - \frac{1}{2} \sum_{\vec{b} > 0}  d(0,\vec{b}) b_i V^i.
\end{equation}
We should note that when $s = 1$, 
the sum of $ \kappa_a y_2^a $ part and the polylogarithm sum term is, up to an overall constant,
$\log |\Phi_\psi(T,U, V)|$, where $\Phi_\psi(T,U, V)$ is a Siegel form which is the exponential or Borcherds lift
of the weight zero Jacobi form $\psi_{0}$ \cite{borcherds1995automorphic,Gritsenko:1996tm}.

We can now use these results for $ \Delta_\text{gauge}$ and  $  \Delta_\text{grav}$ in one-loop expressions
for $g_{\text{gauge}}$ and $g_{\text{grav}}$ (\eqref{eq:string_gauge} and \eqref{eq:string_grav}), and then
compare them with the field theoretical 
expressions \cite{Kaplunovsky:1995jw, deWit:1995zg}
\begin{equation}\label{eq:field_gauge}
 \frac{1}{g_\text{gauge}^2 (p^2)} = \Re \lp -\ii \tilde{S} - \frac{1}{2(s+4)\pi^2} \log \Psi_\text{gauge} \rp + 
     \frac{b_\text{gauge}}{\fpi} \lp \log \frac{M_\text{Planck}^2}{p^2} + \text{K} \rp,
\end{equation}
and
\begin{equation}\label{eq:field_grav}
 \frac{1}{g_\text{grav}^2 (p^2)} = \Re \lp F_1^{het} \rp + 
     \frac{b_\text{grav}}{\fpi} \lp \log \frac{M_\text{Planck}^2}{p^2} + \text{K} \rp.
\end{equation}
Here, $\text{K}$ is the K\"{a}hler potential,
\begin{equation}
 \text{K} = - \log \Re \lp -\ii S \rp - \log\lp-(y_2,y_2)\rp + \text{const}.
\end{equation}
Planck scale, $M_\text{Planck}$, is related to the string scale, $M_\text{str}$, by
\begin{equation}
 M_\text{Planck}^2 = M_\text{str}^2 \Re \lp -\ii S \rp.
\end{equation}
Finally, $\tilde{S}$ and $\Delta^\text{univ}$ are determined through the one-loop contribution to the 
prepotential, $\mc{F}_0^{(1)}$, as
\begin{equation}
 \frac{\Delta^\text{univ}}{\fpi} = \frac{1}{-(y_2,y_2)} \Re \lp \mc{F}_0^{(1)} 
						- \ii y_2^a \frac{\del}{\del y^a} \mc{F}_0^{(1)} \rp,
\end{equation}
and
\begin{equation}
 - \ii \tilde{S} = - \ii S - \frac{1}{s+4} \eta^{a b}  \frac{\del}{\del y^a}  \frac{\del}{\del y^b} \mc{F}_0^{(1)}.
\end{equation}

Setting $ \frac{1}{2(s+4)\pi^2} \log \Psi_\text{gauge} + \frac{b_\text{gauge}}{\fpi} \log\lp-(y_2,y_2)\rp$
equal to the T-duality invariant quantity $\frac{1}{\fpi} \lp \frac{1}{s+4} \nabla^2 -1 \rp \Delta_\text{gauge}$
as in \cite{Henningson:1996jz}, one gets a differential equation for $\mc{F}_0^{(1)}$, which can be solved by
\begin{equation}
 \mc{F}_0^{(1)} = - \frac{\ii}{4 \pi} \frac{1}{6}  d_{a b c} y^a y^b y^c + \frac{1}{64 \pi^4} c(0) \zeta(3)
		  + \frac{1}{32 \pi^4} \sum_{r>0} c(r) \Li_3 \lp e^{2 \pi \ii r.y} \rp,
\end{equation}
which together with the tree level contribution 
\begin{equation}
 \mc{F}_0^{(0)} = \frac{\ii}{2} S (y, y)  =- \ii S (T U - \frac{1}{2} V^i d_{ij} V^j),
\end{equation}
gives the prepotential, $\mc{F}_0^{het}$, at the perturbative level (as further contributions
are non-perturbative). The Laplacian is given by
\begin{equation}
 \nabla^2 = - 2 (y_2, y_2) \lp \eta^{a b} - \frac{2}{(y_2,y_2)} y_2^a y_2^b  \rp \del_a \bar{\del}_b.
\end{equation}

However, the
differential equation has homogeneous solutions as well, and in particular, the cubic terms can be changed by terms of the form
$-\ii \tilde{\rho}_a y^a (y,y)$ where $\tilde{\rho}_a$ are arbitrary real coefficients. This corresponds to the fact
that dilaton, $S$, can be shifted by linear terms in $y^a$ (as discussed in \cite{Louis:1996mt}) without having physical
consequences. We will fix the form appearing in the equation above by requiring no $TU$ factor to appear among the cubic terms.

With the prepotential found in this way, we can also determine the Wilsonian gravitational coupling, $F_1^{het}$, as
\begin{equation}
   F_1^{het} = 24 \lp - \ii S \rp - \frac{\ii}{4 \pi} \rho_a y^a + \frac{1}{4 \pi^2}  
		\sum_{r>0} \tilde{c}(r) \Li_1 \lp e^{2 \pi \ii r.y} \rp + \text{const}.
\end{equation}

Lastly, we note that gravitational coupling constant can be written in the form
\begin{align}\label{eq:siegel2}
 \frac{1}{g_\text{grav}^2 (p^2)} =  24 & \Re \lp -\ii \tilde{S} \rp 
	  +  \frac{b_\text{grav}}{\fpi} \lp \log \frac{M_\text{str}^2}{p^2} - \log\lp -(y_2,y_2) \rp \rp
	  + \text{const} \notag\\
	  &+ \frac{6}{(s+4)\pi^2}  \Bigg[
		 \sum_{r>0} \lp -\frac{r_a r^a}{2} c(r)  +\frac{s+4}{24} \tilde{c}(r) \rp 
		 \Re \lp \Li_1 \lp e^{2 \pi \ii r.y} \rp \rp \notag \\
		 &+  2 \pi \lp \frac{1}{2}d^a_{\ a e} y_2^e +\frac{s+4}{48} \rho_a y_2^a  \rp
	  \Bigg].
\end{align}
Working out the linear term in the brackets and using the fact that $\phi_{-2, \lt} = -\frac{2}{q} + O(1)$ as 
$\T \rightarrow \ii \infty$, it is easy to verify that the term in the bracket comes from the exponential lift
(in particular, for $s=1$ case it is of the form $\log |\Phi_\phi(T,U, V)|$ where $\Phi_\phi(T,U,V)$ is a Siegel form)
of a weight zero Jacobi form with Fourier coefficients 
\begin{equation}
 -\frac{r_a r^a}{2} c(r)  +\frac{s+4}{24} \tilde{c}(r).
\end{equation}
This weight zero Jacobi form can be obtained by
$\mc{L}_{-2} \phi_{-2, \lt}$, where $ \mc{L}_{k}$ is the modified
heat operator
\begin{equation}\label{eq:JacobiDiff}
 \mc{L}_{k} = q \frac{\del}{\del q} -\frac{1}{2} \sum_{a,b} \lp y^a \frac{\del}{\del y^a} \rp 
				  \eta^{a b} \lp y^b \frac{\del}{\del y^b} \rp + \frac{s-2k}{24} E_2,
\end{equation}
mapping a weight $k$ Jacobi form to a weight $k+2$ Jacobi form \cite{Kawai:1998md, Zagier}.
\footnote{This way of writing the coupling constants in terms of exponential lifts $\Phi_\psi$ and
$\Phi_\phi$ is in accordance with the results of section 4 in \cite{Henningson:1996jz}. }

\subsubsection*{Weyl Chambers}
In \cite{Borcherds:1996automorphic} a generalization of the notion of Weyl chambers is used.
Weyl chambers are defined as components
of the $\{ y^a \}$ space where integrals such as those involved in
\eqref{eq:gaugethre} and \eqref{eq:gravthre} are real analytic. In particular, equations \eqref{eq:int_grav} and
\eqref{eq:int_gauge} have terms such as $\sum_{r>0} c(r) \Re \lb \Li_n (x) \rb$ which are obtained by replacing the
infinite series expansion of polylogarithms by $\Li_n$ functions. This is valid if the variable, $x$, has modulus less than 
unity. In other words, expressions in \eqref{eq:int_grav} and
\eqref{eq:int_gauge} are valid if the condition
\begin{equation}
 r.y_2 > 0
\end{equation}
is satisfied for every $r>0$ with $c(r) \neq 0$. We should note that expressions in \eqref{eq:int_grav},
\eqref{eq:int_gauge} and Weyl Chambers associated with
them depend on the ordering introduced on the lattice $U \otimes \lt^*$. The cubic terms in the integral (again suppose
that we eliminate $TU$ terms along with quadratic and quartic ambiguities
as discussed in \cite{Harvey:1995fq}) 
change when one crosses from one Weyl chamber to another. This change can also be computed
by analytically continuing the polylogarithms across the boundaries of the regions introduced above.

Now, we will try to characterize a set of simpler conditions on $y_2$ so that it is ensured that integrals are real analytic on 
regions satisfying those simpler conditions. A piece of information we use is that Fourier coefficients
involved in the integrals are zero unless $r.r = -2 k l + b_i d^{ij} b_j \leq 2$. Therefore, we will be looking for $s+2$
vectors $r_\mu$ where $\mu = -1, 0, \ldots, s$ such that $r_\mu . r_\mu \leq 2 $ and $r_\mu > 0$ for all $\mu$ and moreover
every vector $r$ satisfying $r.r \leq 2$ and $r > 0$ can be decomposed as
\begin{equation}
 r = a^\mu r_\mu,
\end{equation}
where $a^\mu$'s are $s+2$ nonnegative integers. The problem of existence and
uniqueness of such a basis (given an ordering) is a generalization of similar problems studied in
\cite{Harvey:1995fq} and \cite{Hohenegger:2011ff}.

We will start by supposing that the finite set 
\begin{equation}
 \mc{S}_\lt = \{ \vec{b} \in \lt^* | \vec{b}^2 \leq 2  \text{ and } \vec{b} > 0 \}
\end{equation}
has elements $\vec{\varpi}_i \in \mc{S}_\lt$ ($i = 1, 2, \ldots, s$) such that they create the elements of
$\mc{S}_\lt$ with nonnegative coefficients and they form an integral basis for $\lt^*$. Furthermore, the components
of the most positive element $\vec{\theta}$ of $\mc{S}_\lt$, $\vec{\theta}[i]$, should satisfy
\begin{equation}
 \vec{\theta}[i] \geq \vec{b}[i]
 \end{equation}
for all elements $\vec{b}$ in $\mc{S}_\lt$.
Such a basis is unique if it exists and it is possible to construct 
it as follows. First, one should put the positive elements of $\mc{S}_\lt $ in increasing order. Then,
starting with the smallest element, one inductively adds an element in this list to the basis set if it increases
the rank of the vectors chosen at this point by one. At the end, one should check whether the conditions above are satisfied.

We start the construction of $r_\mu$ basis by noting that vectors, $(0,1,\vec{b})$ and
$(0,0,\vec{b})$ for $\vec{b} \in  \mc{S}_\lt$, satisfy $r.r \leq 2$ and $r>0$. Since such vectors span a $s+1$ dimensional
space over reals and since no positive vector $r$ can have negative $k$, there is only one basis vector 
with nonnegative $k$,
which we will choose to be $r_{-1}$. Moreover, vectors  $(0,1,\vec{b})$ and $(0,0,\vec{b})$ are spanned by
$r_0, r_1, \ldots, r_s$. Similarly, since vectors $(0,0,\vec{b})$ span a $s$ dimensional space over reals and since all of
$r_0, r_1, \ldots, r_s$ has $k = 0$, only one of them can have nonzero $l$. Let us choose this basis vector to be $r_0$.
Then, the uniqueness of $\vec{\varpi}_i $ basis gives
\begin{equation}
 r_i = (0, 0, \vec{\varpi}_i) \text{ for } i = 1, 2, \ldots, s.
\end{equation}

Next, from the fact that vectors $r=(0,1, -\vec{b})$ satisfy $r.r \leq 2$ and $r>0$, and the requirement 
that such vectors should be
spanned by $r_0, r_1, \ldots, r_s$ with nonnegative coefficients, $r_0$ is fixed to be
\begin{equation}
 r_0 = (0, 1, -\vec{\theta}).
\end{equation}
Finally, since there are vectors $r$ satisfying our conditions with negative $l$, $r_{-1}$ should have negative $l$ and this
is only possible if we fix
\begin{equation}
 r_{-1} = (1, -1, \vec{0}).
\end{equation}

At this point, the $r_\mu$'s chosen so far span vectors $r$ satisfying $r.r \leq 2$, $r > 0$, and $kl = -1 \text{ or } 0$, with 
positive integer
coefficients. We can prove that this is also valid for vectors with $k l > 0$ provided that 
\begin{equation}
 (\vec{\xi}_i . \vec{\theta})^2 \geq (\vec{\xi}_i)^2,
\end{equation}
where $\vec{\xi}_i$ are vectors satisfying $\vec{\xi}_i . \vec{\varpi}_j = \delta_{i j}$. 
We can prove this assertion as follows.
Suppose $r = ( k, l, \vec{b})$, where $k,l>0$ and $| \vec{b} | ^2 \leq 2(kl+1) $.
Since
\begin{equation}
 r = k r_{-1} + (k+l) r_0 + (0,0, \vec{b} + (k+l) \vec{\theta}),
\end{equation}
we should be able to write the vector $a_i \vec{\varpi}_i \equiv \vec{b} + (k+l) \vec{\theta}$
in $\vec{\varpi}_i$ basis with 
nonnegative integer coefficients ($a_i \geq 0$). To pick the coefficients $a_j$, we multiply this vector with $\vec{\xi}_j$.
\begin{equation}
 a_j = \vec{\xi}_j . \vec{b} + (k+l) \vec{\xi}_j.\vec{\theta}.
\end{equation}
By the Cauchy-Schwarz inequality 
\begin{equation}
 |\vec{\xi}_j . \vec{b} | \leq \sqrt{|\vec{\xi}_j|^2 |\vec{b}|^2} \leq |\vec{\xi}_j| \sqrt{2kl +2}.
\end{equation}
Therefore,
\begin{equation}
 a_j \geq (k+l)  \vec{\xi}_j.\vec{\theta} - |\vec{\xi}_j| \sqrt{2kl +2},
\end{equation}
which can be proved to be nonnegative for $k,l>0$ given that $(\vec{\xi}_i . \vec{\theta})^2 \geq (\vec{\xi}_i)^2$.

Finally, having found these basis vectors, we can assert that the integral expressions in threshold corrections are valid
in the chamber
\begin{equation}
 r_\mu . y_2 > 0 \text{ for all } \mu = -1, 0, \ldots, s.
\end{equation}
Note that sums of the form $\sum_{r>0} c(r) \Li_n(.)$ can be rewritten as $\sum_{a_\mu \in N^{s+2}} c(a_\mu r^\mu) \Li_n(.)$
if we introduce the set $N^{s+2} \equiv \lp \ZZ^+_0 \rp^{s+2} - \{ (0,\ldots,0) \}$.

\subsubsection*{Example 1 : $\mathbf{\lt = \langle 2 m \rangle}$ with $\mathbf{m \in \ZZ^+}$}
If $\lt = \langle 2 m \rangle$, the dual lattice is $\lt^* = \langle 1/2 m \rangle$. Let us take a generator of $\lt^*$,
$\vec{\gamma}$, and denote any vector in $\lt^*$ by its component with respect to $\vec{\gamma}$. In other words,
we take the generator to be the vector $\vec{\gamma} = (1)$ where it satisfies $(1).(1) = 1/2m$. Vector $(1)$
also generates the set, $\mc{S}_\lt$, which is the set of all positive vectors in $\lt^*$ with norm-square less than or 
equal to two. So, we also set $\vec{\varpi} = (1)$, compute the 
highest vector as $\vec{\theta} = (\lfloor \sqrt{4m} \rfloor)$ and the dual basis vector as $\vec{\xi} = (2m)$.
Construction discussed in this section gives
\begin{align}
 r_{-1} &= (1,-1,0), \\
 r_{0} &= (0,1,-\lfloor \sqrt{4m} \rfloor), \\
 r_{1} &= (0,0,1). 
\end{align}
Now, we can check that the last requirement 
\begin{equation}
\vec{\xi} . \vec{\theta} = \lfloor \sqrt{4m} \rfloor \geq |\vec{\xi}| = \sqrt{2m}
\end{equation}
is satisfied for all positive integers m.

In particular, for the $m=2$ example discussed above,
the cubic part of the prepotential and the linear part of the gravitational coupling
is given by the following expression in the chamber $\Im T > \Im U > 2 \Im V^1 > 0$:
\begin{align}
 F_0^{het,cub} &\equiv -4 \pi S \frac{(y,y)}{2} + \frac{1}{6}  d_{a b c} y^a y^b y^c \\
		&= 4 \pi S \left[T U-2 (V^1)^2\right]-2 T (V^1)^2+\frac{U^3}{3}-4 U (V^1)^2+8 (V^1)^3,
\end{align}
and
\begin{align}
  F_1^{het,lin} &\equiv 24 (4 \pi S) + \rho_a y^a \\
	      &= 24 (4 \pi S) + 24 T+44 U-96 V^1.
\end{align}

For our $m=3$ example, we have the following quantities in the chamber $\Im T > \Im U > 3 \Im V^1 > 0$:
\begin{equation}
 F_0^{het,cub} = 4 \pi S \left[T U-3 (V^1)^2\right]-3 T (V^1)^2+\frac{U^3}{3}-6 U (V^1)^2+14 (V^1)^3,
\end{equation}
and
\begin{equation}
  F_1^{het,lin} = 24 (4 \pi S) + 24 T+44 U-132 V^1.
\end{equation}

\subsubsection*{Example 2 : $\mathbf{\lt = A_2}$}
For the $\lt = A_2$ model discussed previously, we can go through the same exercise and find
\begin{align}
 F_0^{het,cub} = 4 \pi S & \left[T U-(V^1)^2+V^2 V^1-(V^2)^2\right]-T (V^1)^2-T (V^2)^2+T V^1 V^2+\frac{U^3}{3} \notag\\
 & -2 U (V^1)^2  -2 U (V^2)^2 
    +2 U V^1 V^2+\frac{10 (V^1)^3}{3}+\frac{4 (V^2)^3}{3} \notag\\ 
    &+4 V^1 (V^2)^2-5 (V^1)^2 V^2,
\end{align}
and
\begin{equation}
  F_1^{het,lin} = 24 (4 \pi S) + 24 T+44 U-52 V^1-4 V^2,
\end{equation}
in the chamber
\begin{equation}
 \Im \lp T - U \rp, \Im \lp U-2 V^1+V^2 \rp,\Im \lp V^1-2 V^2 \rp,\Im \lp V^2 \rp > 0.
\end{equation}

\subsubsection*{Example 3 : $\mathbf{\lt = A_3}$}
For our model with $\lt = A_3$, we have
\begin{align}
 F_0^{het,cub} = 4 \pi S & \left[T U-(V^1)^2+V^2 V^1-(V^2)^2-(V^3)^2+V^2 V^3\right]-T (V^1)^2-T (V^2)^2 \notag\\ 
  &-T (V^3)^2+T V^1 V^2+T V^2 V^3+\frac{U^3}{3}-2 U (V^1)^2-2 U (V^2)^2-2 U (V^3)^2 \notag\\
  &+2 U V^1 V^2+2 U V^2 V^3+\frac{10 (V^1)^3}{3}+\frac{4 (V^2)^3}{3}+\frac{4 (V^3)^3}{3} +4 V^1 (V^2)^2\notag\\
  &+2 V^1 (V^3)^2+3 V^2 (V^3)^2-5 (V^1)^2 V^2-4 (V^2)^2 V^3-2 V^1 V^2 V^3,
\end{align}
and
\begin{equation}
  F_1^{het,lin} = 24 (4 \pi S) + 24 T+44 U-52 V^1-4 V^2-4 V^3,
\end{equation}
in the chamber
\begin{equation}
 \Im \lp T - U \rp, \Im \lp U-2 V^1+V^2 \rp,\Im \lp V^1-2 V^2+V^3 \rp,\Im \lp V^2-2 V^3 \rp, \Im \lp V^3 \rp > 0.
\end{equation}

\subsection{Models with a Single Wilson Line and Paramodular Groups}
In this section, we will gather details on the $\lt = \langle 2m \rangle$ case and discuss their relation to paramodular groups. 
Pieces of such details have already appeared in the preceding sections; however, it will be useful to put these 
together both because $s=1$ case is the simplest example of $\mc{N}=2$ heterotic compactifications with Wilson lines
that also has partially appeared in the literature and because studying the implications of the T-duality group,
which is a paramodular group in this case, on a 
possible Type-II dual partner would be easier for this relatively simple case.

The classical vector multiplet moduli space for this case is $SO(3,2)/SO(3)\otimes SO(2)$ when we separate the factor coming from
the axio-dilaton.  This space can be parametrized by three complex moduli $T$, $U$, and $V$. To compare with the previous
sections we set $y = (T, U, V \vec{\xi})$ where $\vec{\xi}$ is a generator for the lattice $\lt = \langle 2m \rangle$,
in other words $\vec{\xi}.\vec{\xi} = 2 m$. State charges, on the other hand, will be of the form 
$(m_1, m_2, n_1, n_2, b \vec{\gamma}) \in \lp U \oplus U \oplus \lt^* \rp$, where $\vec{\gamma}$ 
is a generator of the dual lattice
$\lt^* = \langle 1/2m \rangle$, and hence charges can be parametrized by five integers $m_1$, $m_2$, $n_1$, $n_2$ and $b$.
Then, automorphisms of this lattice which preserve the norm
\begin{equation}
 \frac{ p_L^2 - p_R^2}{2} = \frac{b^2}{2} \vec{\gamma}.\vec{\gamma} - m_1 n_1 + m_2 n_2 = \frac{b^2}{4m} - m_1 n_1 + m_2 n_2 ,
\end{equation}
as well as the conjugacy class in $\lt^* / \lt$ (or in this case $b(\text{mod }2m)$)
induce T-duality transformations together with the transformation changing the sign of all charges.
We can generate these transformations from four basic operations as explained in the
subsequent paragraphs \cite{Neumann:1996is}.

The first two are generated by the automorphisms
\begin{equation}
 m_1 \rightarrow m_1 + n_2 \text{ and } m_2 \rightarrow n_1 + m_2,
\end{equation}
and
\begin{equation}
 m_2 \rightarrow -n_1, n_1 \rightarrow m_2, m_1 \rightarrow -n_2, \text{ and } n_2 \rightarrow m_1
\end{equation}
which induce
\begin{equation}\label{eq:Tdual1.1}
 T \rightarrow T+1,
\end{equation}
and
\begin{equation}\label{eq:Tdual1.2}
 T \rightarrow -\frac{1}{T}, U \rightarrow U - m \frac{V^2}{T}, V \rightarrow \frac{V}{T},
\end{equation}
respectively.
We see that these two transformations together generate a subgroup $SL(2,\ZZ)_T$ which acts on the moduli as
\begin{equation}\label{eq:Tdual1}
 T \rightarrow \frac{aT+b}{cT+d}, U \rightarrow U - \frac{mcV^2}{cT+d}, V \rightarrow \frac{V}{cT+d},
\end{equation}
where $a,b,c,d$ are integers satisfying $ad - bc = 1$. Note that by picking $a=d=-1$ and $b=c=0$ one gets the transformation
$V \rightarrow -V$.

The next T-duality transformation is induced by the automorphism
\begin{align}
 m_2 &\rightarrow m_2 - \alpha^2 m n_2 + \alpha b, \notag\\
 n_1  &\rightarrow n_1 + \lambda^2 m m_1 - 2 \lambda \alpha m n_2 + \lambda b \notag\\
 b &\rightarrow b + 2 \lambda m m_1 - 2 \alpha m n_2,
\end{align}
where $\lambda, \alpha \in \ZZ$,
and is given by
\begin{align}\label{eq:Tdual2}
 T &\rightarrow T \notag\\
 U &\rightarrow U + \lambda^2 m T + 2 \lambda m V \notag\\
 V &\rightarrow V + \lambda T + \alpha.
\end{align}

The final transformation is induced by
\begin{equation}
 m_1 \leftrightarrow n_1
\end{equation}
which gives
\begin{equation}\label{eq:Tdual3}
 T \leftrightarrow U.
\end{equation}

The threshold corrections, then, involve automorphic forms of the T-duality transformations induced by \eqref{eq:Tdual1},
\eqref{eq:Tdual2} and \eqref{eq:Tdual3}. In particular, equations \eqref{eq:int_gauge} and \eqref{eq:siegel2} involve
Siegel forms $\Phi_\psi(Z)$ and $\Phi_\phi(Z)$ which are functions on the genus 2 Siegel upper half plane $\mc{H}_2$,
where we parametrize the elements $Z$ as
\begin{equation}
 Z = \left( \begin{array}{cc} p & q \\ q & r \end{array} \right) \equiv 
    \left( \begin{array}{cc} T & V \\ V & U/m \end{array} \right) \in \mc{H}_2,
\end{equation}
for $\Im p > 0$, $\Im r > 0$ and $\Im \det Z > 0$. These two functions are Siegel forms over a discrete subgroup of
$Sp(4,\RR)$ which can be identified with the T-duality group up to an overall sign. This discrete subgroup is a particular
semi-direct product of $\ZZ_2$ with $\Gamma_m$ for $m > 1$, where $\Gamma_m$ is a paramodular group.
We will call this semi-direct product an extended paramodular group and denote it by $\Gamma_m^+$.
For $m=1$, the action of
$\ZZ_2$ (which comes from the $T \leftrightarrow U$ exchange symmetry) is already included in $\Gamma_1$ which is isomorphic
to $Sp(4,\ZZ)$. Further details on paramodular groups can be found in \cite{neumann1998algebras} and \cite{Gritsenko:1996tm}.
Note that $\mc{H}_2 / \Gamma_m^+$ is isomorphic to the moduli space of K3 surfaces with polarization type 
$2 E_8(-1) \oplus \langle 2m \rangle$ \cite{gritsenko1998minimal}.

We can now study the
particular heterotic models with $m = 1, \ldots, 5$ we studied in the previous sections. In equation
\eqref{eq:siegel2} it was noted that (for $s=1$)
\begin{align}
 \frac{1}{g_\text{grav}^2 (p^2)} =  24 & \Re \lp -\ii \tilde{S} \rp 
	  +  \frac{\tilde{c}(0)}{\fpi} \lp \log \frac{M_\text{str}^2}{p^2} - \log\lp -(y_2,y_2) \rp \rp
	  + \text{const} \notag\\
	  &+ \frac{1}{10 \pi^2}  \log | \Phi_m (Z) |,
\end{align}
where the Siegel form, $\Phi_m (Z)$ is the exponential lift\footnote{See theorem 2.1 of \cite{Gritsenko:1996tm}.} of 
the weight zero, index $m$, nearly holomorphic Jacobi form
\begin{equation}
 \chi_{0,m} \equiv 12 \mc{L}_{-2} \phi_{-2,m}.
\end{equation}
From physical requirements it is easy to see that the Fourier coefficients of $\chi_{0,m}$, which we will denote as $f(n,r)$,
are integral. We will also use the notation $\Phi_m = \text{Exp-Lift}(\chi_{0,m})$ for exponential lift.

It is also found in \cite{Gritsenko:1996tm} that the divisors of $\Phi_m$ on $\mc{H}_2 / \Gamma_m^+$ are Humbert surfaces
\begin{equation}
 H_D(b) = \pi_m \lp \{ Z \in \mc{H}_2 | a T + b V + U = 0 \}   \rp,
\end{equation}
where $\pi_m$ projects from $\mc{H}_2$ to $\mc{H}_2 / \Gamma_m^+$, the
discriminant $D > 0$ is defined as $D = b^2 - 4 m a$, $b$ can be restricted to particular representatives
of $\pm b(\text{mod }2m)$ and 
divisor multiplicities are given by
\begin{equation}
 m_{D,b} = \sum_{n>0} f(n^2 a, n b).
\end{equation}
Physically, these are T-duality inequivalent surfaces on the vector multiplet moduli space on which there are BPS states that become massless
hence creating a singularity in $\log \Phi_m$.

\subsubsection*{$\mathbf{m=1}$ Example}
From equation \eqref{eq:m1} we can compute $\chi_{0,1}$ as
\begin{align}
 \chi_{0,1} = &\frac{19}{q}+\left(19 y^2-28
   y+1050-\frac{28}{y}+\frac{19}{y^2}\right) \notag\\
   &+\left(1050 y^2+617088
   y+2504520+\frac{617088}{y}+\frac{1050}{y^2}\right) q + \ldots.
\end{align}
The Siegel form it lifts to has its divisor as $19 H_4 - 9 H_1$. In the notation of \cite{Gritsenko:1996tm} and 
\cite{gritsenko1996igusa}, $\Delta_{30}(Z)$ and $\Delta_{5}(Z)$ are Siegel forms of $Sp(4,\ZZ)$ with divisors
$H_4$ and $H_1$, respectively. This gives
\begin{equation}
 \log  \Phi_1  = 19 \log \Delta_{30} - 9 \log \Delta_{5} + \text{const}.
\end{equation}
Moreover $\Delta_{30} = \text{Exp-Lift}(\rho_{0,1})$ where
\begin{equation}
 \rho_{0,1} = \frac{1}{q}+\left( y^2- y+ 60 -\frac{1}{y}+\frac{1}{y^2}\right) + \ldots,
\end{equation}
and $\Delta_{5} = \text{Exp-Lift}(\kappa^1_{0,1})$ where
\begin{equation}
 \kappa^1_{0,1} = \tilde{\phi}_{0,1} 
	  = \left(y+10+\frac{1}{y}\right)+\frac{2 (y-1)^2 \left(5 y^2-22 y+5\right) q}{y^2} + \ldots.
\end{equation}
This finally implies that
\begin{equation}
 12 \mc{L}_{-2} \phi_{-2,1} = 19 \rho_{0,1} - 9 \kappa^1_{0,1}.
\end{equation}
This relation was first noticed in the context of threshold corrections in \cite{LopesCardoso:1996nc}.

\subsubsection*{$\mathbf{m=2}$ Example}
Equation \eqref{eq:m2} gives $\chi_{0,2}$ as
\begin{align}
 \chi_{0,2} = &\frac{19}{q}+\left(96 y+840+\frac{96}{y}\right) 
 +\Big(19 y^4+96 y^3+86632 y^2+894880 y+1777542 \notag\\
  &+ \frac{894880}{y}+\frac{86632}{y^2}+\frac{96}{y^3}+\frac{19}{y^4}\Big) q + \ldots.
\end{align}
The Siegel form it lifts to has its divisor as $19 H_8 + 96 H_1$ on $\mc{H}_2 / \Gamma_2^+$.
\cite{Gritsenko:1996tm} introduces
$\Psi_{12}^{(2)}(Z)$ and $\Delta_{2}(Z)$ which are Siegel forms of $\Gamma_2^+$ with divisors
$H_8$ and $H_1$, respectively. This gives
\begin{equation}
 \log  \Phi_2  = 19 \log \Psi_{12}^{(2)} + 96 \log \Delta_{2} + \text{const}.
\end{equation}
Moreover, $\Psi_{12}^{(2)} = \text{Exp-Lift}(\rho_{0,2})$ where
\begin{equation}
 \rho_{0,2} = \frac{1}{q} + 24 + \ldots,
\end{equation}
and $\Delta_{2} = \text{Exp-Lift}(\kappa^1_{0,2})$ where
\begin{equation}
 \kappa^1_{0,2} = \left(y+4+\frac{1}{y}\right)+ \ldots.
\end{equation}
This finally implies that
\begin{equation}
 12 \mc{L}_{-2} \phi_{-2,2} = 19 \rho_{0,2} + 96 \kappa^1_{0,2}.
\end{equation}
In fact, \cite{LopesCardoso:1996zj} uses exactly this argument 
to find out the complete form of the threshold corrections in $m=2$ case. This argument is closely related to the arguments
we used in the previous sections to match Jacobi forms coming from the index computation.

\subsubsection*{$\mathbf{m=3}$ Example}
Equation \eqref{eq:m3} gives $\chi_{0,3}$ as
\begin{align}
 \chi_{0,3} = &\frac{19}{q}+\left(-9 y^2+180 y+690+\frac{180}{y}-
 \frac{9}{y^2}\right)+\Big(-9 y^4+9768 y^3+212706 y^2+925272 y \notag\\
   &+1445322+\frac{925272}{y}+\frac{212706}{y^2}+\frac{9768}{y^3}-\frac{9}{y^4}\Big) q + \ldots.
\end{align}
The Siegel form it lifts to has its divisor as $19 H_{12} + 171 H_1 - 9 H_4$ on $\mc{H}_2 / \Gamma_3^+$.
\cite{Gritsenko:1996tm} introduces
$\Psi_{12}^{(3)}(Z)$, $\Delta_{1}(Z)$, and $ F_2^{(3)}(Z)$ which are Siegel forms of $\Gamma_3^+$ with divisors
$H_{12}$, $H_1$ and $H_4$, respectively. This gives
\begin{equation}
 \log  \Phi_3  = 19 \log \Psi_{12}^{(3)} + 171 \log \Delta_{1} - 9 \log F_2^{(3)} + \text{const}.
\end{equation}
Furthermore, $\Psi_{12}^{(3)} = \text{Exp-Lift}(\rho_{0,3})$ where
\begin{equation}
 \rho_{0,3} = \frac{1}{q} + 24 + \ldots,
\end{equation}
$\Delta_{1} = \text{Exp-Lift}(\kappa^1_{0,3})$ where
\begin{equation}
 \kappa^1_{0,3} = \left(y+2+\frac{1}{y}\right)+ \ldots,
\end{equation}
and $F_{2}^{(3)} = \text{Exp-Lift}(\kappa^2_{0,3})$ where
\begin{equation}
 \kappa^2_{0,3} = \left(y^2-y+12-\frac{1}{y}+\frac{1}{y^2}\right)+ \ldots.
\end{equation}

This finally implies that
\begin{equation}
 12 \mc{L}_{-2} \phi_{-2,3} = 19 \rho_{0,3} + 171 \kappa^1_{0,3} - 9 \kappa^2_{0,3},
\end{equation}
which can be checked with the explicit expressions of Jacobi forms we have computed.

\subsubsection*{$\mathbf{m=4}$ Example}
Equation \eqref{eq:m4} gives $\chi_{0,4}$ as
\begin{align}
 \chi_{0,4} = &\frac{19}{q}+\left(-8 y^2+224
   y+600+\frac{224}{y}-\frac{8}{y^2}\right)+\Big(570 y^4+38688 y^3+308160y^2+895200y \notag\\
   &+ 1255560+\frac{895200}{y}+\frac{308160}{y^2}+\frac{38688}{y^3}+\frac{570}{y^4}\Big) q + \ldots.
\end{align}
The Siegel form it lifts to has its divisor as $19 H_{16}(0) + 216 H_1 - 8 H_4$ on $\mc{H}_2 / \Gamma_4^+$.
\cite{Gritsenko:1996tm} introduces
$\Psi_{12}^{(4)}(Z)$, $\Delta_{1/2}(Z)$, and $ F_2^{(4)}(Z)$ which are Siegel forms of $\Gamma_4^+$ with divisors
$H_{16}(0)$, $H_1$ and $H_4$, respectively. This gives
\begin{equation}
 \log  \Phi_4  = 19 \log \Psi_{12}^{(4)} + 216 \log \Delta_{1/2} - 8 \log F_2^{(4)} + \text{const}.
\end{equation}
Further, $\Psi_{12}^{(4)} = \text{Exp-Lift}(\rho_{0,4})$ where
\begin{equation}
 \rho_{0,4} = \frac{1}{q} + 24 + \ldots,
\end{equation}
$\Delta_{1/2} = \text{Exp-Lift}(\kappa^1_{0,4})$ where
\begin{equation}
 \kappa^1_{0,4} = \left(y+1+\frac{1}{y}\right)+ \ldots,
\end{equation}
and $F_{2}^{(4)} = \text{Exp-Lift}(\kappa^2_{0,4})$ where
\begin{equation}
 \kappa^2_{0,4} = \left(y^2-y+9-\frac{1}{y}+\frac{1}{y^2}\right)+ \ldots.
\end{equation}

This implies that
\begin{equation}
 12 \mc{L}_{-2} \phi_{-2,4} = 19 \rho_{0,4} + 216 \kappa^1_{0,4} - 8 \kappa^2_{0,4},
\end{equation}
which can again be checked with the explicit expressions of Jacobi forms we have computed.

\subsubsection*{$\mathbf{m=5}$ Example}
Equation \eqref{eq:m5} gives $\chi_{0,5}$ as
\begin{align}
 \chi_{0,5} = &\frac{19}{q}+\left(3 y^2+228 y+570+\frac{228}{y}+\frac{3}{y^2}\right)+\Big(-4 y^5+4802 y^4
 +77532 y^3+368218 y^2 \notag\\  
 &+859048 y +1121604+\frac{859048}{y}+\frac{368218}{y^2}+\frac{77532}{y^3}+\frac{4802}{y^4}-\frac{4}{y^5}\Big) q + \ldots.
\end{align}
For $m=5$, a nearly holomorphic Jacobi form with at most $1/q$ pole can lift to a Siegel form which has the Humbert 
surface $H_5$ with $b=5$ and $a=1$ among its divisors. The multiplicity in the divisor associated with 
this surface is given as
\begin{equation}
m_{5,5} =  g(1, 5) + g(4,10) + g(9,15)
\end{equation}
if $g(n,r)$ are Fourier coefficients of the lifting Jacobi form, which are integers.\footnote{Terms of the form
$g(n^2,5n)$ with $n>3$ are zero and hence do not contribute to $m_{5,5}$. This can be basically proved by the elliptical
transformation property \cite{Zagier}. For a nearly holomorphic Jacobi form with no $q$ pole more severe than $1/q$,
$g(n,r)$ becomes vanishing as soon as $r^2 - 4 n m > m^2 + 4m $.} 
The divisor of the Siegel form it lifts to is $19 H_{20} + 231 H_1 + 3 H_4 + 15 H_5$ on $\mc{H}_2 / \Gamma_5^+$.
In contrast to the previous cases, there are no index $5$ Jacobi forms with a divisor purely on $H_{20}$, $H_1$ or $H_4$. 
However, one can still find Jacobi forms $\rho_{0,5}$, $\kappa^1_{0,5}$ and $\kappa^2_{0,5}$ of the form
\begin{equation}
 \rho_{0,5} = \frac{1}{q} + 24 + \ldots,
\end{equation}
\begin{equation}
 \kappa^1_{0,5} = \left(5 y+2+\frac{5}{y}\right)+ \ldots,
\end{equation}
and
\begin{equation}
 \kappa^2_{0,5} = \left(5 y^2- 5 y+36 -\frac{5}{y}+\frac{5}{y^2}\right)+ \ldots.
\end{equation}
The divisors of the Siegel forms they exponentially lift to are $H_{20} + 3 H_5$, $5 H_1 - H_5$ and $5 H_4 + 7 H_5$. Then,
one concludes that
\begin{equation}
 5 \log  \Phi_5  = 95 \log  \text{Exp-Lift}(\rho_{0,5}) + 231 \log  \text{Exp-Lift}(\kappa^1_{0,5}) 
			  + 3 \log  \text{Exp-Lift}(\kappa^2_{0,5}) + \text{const}.
\end{equation}

This then implies
\begin{equation}
 12 \mc{L}_{-2} \phi_{-2,5} = 19 \rho_{0,5} + \frac{231}{5} \kappa^1_{0,5} + \frac{3}{5} \kappa^2_{0,5},
\end{equation}
which we compare and check with the explicit expressions of Jacobi forms we have computed.

\section{Type IIA - Heterotic String Duality}
In the previous chapter, our discussion was exclusively on $\mc{N} = 2$, $D = 4$ heterotic string models. Now, we can shift our
focus to the dual story for a Type IIA theory compactified on an appropriate Calabi-Yau threefold. First examples
of this duality are discussed in \cite{Kachru:1995wm} and \cite{Ferrara:1995yx}. After the first 
examples, various chains of heterotic models with duals on Calabi-Yau threefolds have been constructed 
\cite{Aldazabal:1995yw, Candelas:1996su, Bershadsky:1996nh, Aldazabal:1996du, Candelas:1997pv}. Our discussion in this
section will be conjectural as compared to the previous sections.

The numbers of vector multiplet and hypermultiplet moduli in a Type IIA compactification
on a Calabi-Yau is determined by the topology of the manifold
as $N_v = h^{1,1}$ and $N_h = h^{2,1} +1$. In particular, if the heterotic models described in the previous chapter
have duals on Calabi-Yau threefolds we can identify
\begin{equation}
 h^{1,1} = s+3 \text{ and } c(0) = 2(N_h - N_v -1) = 2(h^{2,1} - h^{1,1}) = -\chi,
\end{equation}
where $\chi$ is the Euler number associated with the compactification manifold.

Vector moduli are given by expanding the complexified K\"{a}hler form in terms of integral K\"{a}hler
class generators $J_1, \ldots, J_{s+3}$ as
\begin{equation}
 B + \ii J = \sum_i t_i J_i,
\end{equation}
where $B$ is the antisymmetric field, $J$ is the K\"{a}hler form, and vector multiplet moduli satisfy $\Im\lp t_i \rp \geq 0$ in 
the K\"{a}hler cone. 

The prepotential and gravitational coupling are given in terms of the topological properties of the Calabi-Yau (CY) manifold
and hence one can study the duality by comparing the results on heterotic side with this topological data as was done in
\cite{Berglund:1996uy, LopesCardoso:1996nc, Weiss:2007tk, Kawai:1996te, Klemm:2004km}
for models with Wilson lines. To be more concrete, let us give some
definitions.  We define triple intersection numbers as
\begin{equation}
 \kappa_{i j k} = \int J_i \wedge J_j \wedge J_k.
\end{equation}
We will also need $\kappa_i$ which are defined as
\begin{equation}
 \kappa_i = \int J_i \wedge C_2,
\end{equation}
where $C_2$ is the second Chern class of the associated CY threefold.

The prepotential and gravitational couplings for the low energy effective field theory then can be given in terms
of genus-0 and genus-1 Gromov-Witten potentials as in \cite{Aspinwall:1995vk, Kawai:1996te, Kawai:1998md}
\begin{equation}
 F_0^{II} = -\frac{\ii}{6} \sum_{i,j,k} \kappa_{i j k} t_i t_j t_k - \frac{1}{16 \pi^3} \chi \zeta(3)
	      + \frac{1}{8 \pi^3} \sum_{d \in N^{s+3}} N^r(d) \Li_3 \lp e^{2 \pi \ii d.t} \rp,
\end{equation}
and
\begin{equation}
 F_1^{II} = - \ii \pi \sum_{i} \kappa_i t_i + \sum_{d \in N^{s+3}} N^{r,e}(d) \Li_1 \lp e^{2 \pi \ii d.t} \rp.
\end{equation}
Here, $N^r(d)$ are the worldsheet instanton numbers for rational curves of degree $d$. $N^{r,e}(d)$ is 
defined as
\begin{equation}
 N^{r,e}(d) = N^r(d) + 12 \sum_{\substack{d' \in N^{s+3},\\ d' || d}} N^e(d'),
\end{equation}
where $d' || d$ if $d = n d'$ for a positive integer $n$, and $N^e(d)$ is the worldsheet instanton numbers 
for elliptic curves of degree $d$.

From the factorization of the classical moduli space into two parts, one can show that the Calabi-Yau manifold
on the dual side
is a K3 fibration \cite{Aspinwall:1995vk, Aspinwall:1996mn, Aspinwall:2000fd}. 
Moreover, the area of the base of this fibration is controlled by the $SL(2,\RR) / U(1)$ factor in the 
moduli space or in other words by the axio-dilaton, $S$, up to a piece linear in the other vector multiplet moduli. 
Therefore, the perturbative limit $\Im S \rightarrow \infty$ corresponds to the case where the area of the fibration base is
becoming large. Therefore, among the instanton corrections only those having degree $0$ on the base survive. 

In most of the duality papers cited above, the vector multiplet moduli mapping is accomplished by comparing the cubic parts in the 
prepotential. The work of \cite{Kawai:1998md}, however,
conjectures a mapping provided that the lattice $\lt$ is either a root lattice
of a simple Lie algebra or a scaled version of a root lattice by a positive integer.\footnote{This is up to an
ambiguity left in equation 5.27.} In particular, K\"ahler class for the base is mapped as
\begin{equation}
 t_2 = 4 \pi S - U - \frac{n}{2} (T - U).
\end{equation}
This is obtained by assuming that the CY threefold is also an elliptic fibration over the Hirzebruch surface 
${\mathbb F}_n$. On the heterotic side, $n$ can be read from the distribution of instantons
$(12 + n, 12 - n)$ if it is a smooth compactification. In the following, we will restrict to the models that can be obtained
starting from the $SU(6)$ model in \eqref{eq:SU6model}. Since this can also be obtained in a smooth
compactification with $n=2$,
we will take $n=2$ in the examples we discuss below.

The form of $F_0^{II}$ and $F_1^{II}$ suggests that we identify them with the perturbative heterotic results as
\begin{equation}
 F_0^{II}|_{\Im t_2 \rightarrow \infty} = 4 \pi \mc{F}_0^{het} \text{ and } 
		    F_1^{II}|_{\Im t_2 \rightarrow \infty} = 4 \pi^2   F_1^{het},
\end{equation}
and that we identify vector multiplet moduli as
\begin{align}
 t_1 &= r_0 . y = U - \vec{\theta}.(V^j \vec{\beta_j}), \\
 t_3 &= r_{-1}.y = T - U, \\
 t_{i+3} &= r_i.y = \vec{\varpi}_i.(V^j \vec{\beta_j}) \text{ for } i = 1,2, \ldots, s.
\end{align}

In the following, we will use these mappings
for some of the heterotic models we studied.

\subsubsection*{Example 1 : $\mathbf{\lt = \langle 4 \rangle}$}
\begin{equation}
 t_1\to U-2 V^1,t_2\to S-T,t_3\to T-U,t_4\to V^1,
\end{equation}
\begin{equation}
 \sum_{i} \kappa_i t_i = 92 t_1+24 t_2+48 t_3+88 t_4,
\end{equation}
\begin{align}
 \frac{1}{6} \sum_{i,j,k} \kappa_{i j k} t_i t_j t_k = 
	  &\frac{4 t_1^3}{3}+t_2 t_1^2+2 t_3 t_1^2+8 t_4 t_1^2+t_3^2 t_1+8 t_4^2 t_1+t_2 t_3 t_1+4 t_2 t_4 t_1 \notag\\
	  &+8 t_3 t_4 t_1+\frac{8 t_4^3}{3}+2 t_2 t_4^2+4 t_3 t_4^2+2 t_3^2 t_4+2 t_2 t_3 t_4.
\end{align}
One can now read the instanton numbers $N^r(d)$ and $N^e(d)$ using the weight -2 Jacobi form $\phi_{-2,2}$.
This example was previously studied in \cite{Kawai:1998md}.

\subsubsection*{Example 2 : $\mathbf{\lt = \langle 6 \rangle}$}
\begin{equation}
 t_1\to U-3 V^1,t_2\to S-T,t_3\to T-U,t_4\to V^1,
\end{equation}
\begin{equation}
 \sum_{i} \kappa_i t_i = 92 t_1+24 t_2+48 t_3+144 t_4,
\end{equation}
\begin{align}
 \frac{1}{6} \sum_{i,j,k} \kappa_{i j k} t_i t_j t_k = 
	  &\frac{4 t_1^3}{3}+t_2 t_1^2+2 t_3 t_1^2+12 t_4 t_1^2+t_3^2 t_1+24 t_4^2 t_1+t_2 t_3 t_1+ 6 t_2 t_4 t_1 \notag\\
	 & +12 t_3 t_4 t_1+14 t_4^3+6 t_2 t_4^2+12 t_3 t_4^2+3 t_3^2 t_4+3 t_2 t_3 t_4.
\end{align}

\subsubsection*{Example 3 : $\mathbf{\lt = A_2}$}
\begin{equation}
 t_1\to U-2 V^1+V^2,t_2\to S-T,t_3\to T-U,t_4\to V^1-2 V^2,t_5\to V^2,
\end{equation}
\begin{equation}
 \sum_{i} \kappa_i t_i = 92 t_1+24 t_2+48 t_3+132 t_4+168 t_5,
\end{equation}
\begin{align}
 \frac{1}{6} \sum_{i,j,k} \kappa_{i j k} t_i t_j t_k = 
	  &\frac{4 t_1^3}{3}+t_2 t_1^2+2 t_3 t_1^2+8 t_4 t_1^2+12 t_5 t_1^2+t_3^2 t_1+12 t_4^2 t_1+24 t_5^2 t_1 \notag\\
	  &+t_2 t_3 t_1+4 t_2 t_4 t_1+8 t_3 t_4 t_1+6 t_2 t_5 t_1+12 t_3 t_5 t_1+36 t_4 t_5 t_1+6 t_4^3+16 t_5^3 \notag\\
	  &+3 t_2 t_4^2+6 t_3t_4^2+6 t_2 t_5^2+12 t_3 t_5^2+36 t_4 t_5^2+2 t_3^2 t_4+2 t_2 t_3 t_4+3 t_3^2 t_5\notag\\
	  &+27 t_4^2 t_5+3 t_2 t_3 t_5+9 t_2 t_4 t_5+18 t_3 t_4 t_5.
\end{align}

Intersection numbers for $\mathbb{P}^4(1,1,2,6,8)[18]$, which has Hodge numbers $(5,161)$, is given in \cite{Berglund:1996uy}
matching the result above. Heterotic computation at an orbifold point and moduli mapping is also given in \cite{Weiss:2007tk},
where the map was found by comparison to topological information given in \cite{Berglund:1996uy}.

\subsubsection*{Example 4 : $\mathbf{\lt = A_3}$}
\begin{equation}
 t_1\to U-2 V^1+V^2,t_2\to S-T,t_3\to T-U,t_4\to V^1-2 V^2+V^3,t_5\to V^2-2 V^3,t_6\to V^3,
\end{equation}
\begin{equation}
 \sum_{i} \kappa_i t_i = 92 t_1+24 t_2+48 t_3+132 t_4+168 t_5+200 t_6,
\end{equation}
\begin{align}
 \frac{1}{6} \sum_{i,j,k} \kappa_{i j k} t_i t_j t_k = 
	  &\frac{4 t_1^3}{3}+t_2 t_1^2+2 t_3 t_1^2+8 t_4 t_1^2+12 t_5 t_1^2+16 t_6 t_1^2+t_3^2 t_1
	  +12 t_4^2 t_1+24 t_5^2 t_1\notag\\
	  &+40 t_6^2 t_1+t_2 t_3 t_1+4 t_2 t_4 t_1+8 t_3 t_4 t_1+6 t_2 t_5 t_1+12 t_3 t_5 t_1
	  +36 t_4 t_5 t_1\notag\\
	  &+8 t_2 t_6 t_1+16 t_3 t_6 t_1+48 t_4 t_6 t_1+64 t_5 t_6 t_1+6 t_4^3+16 t_5^3+\frac{100 t_6^3}{3}
	  +3t_2 t_4^2\notag\\
	  &+6 t_3 t_4^2+6 t_2 t_5^2+12 t_3 t_5^2+36 t_4 t_5^2+10 t_2 t_6^2+20 t_3 t_6^2+60 t_4 t_6^2+80 t_5 t_6^2\notag\\
	  & +2 t_3^2 t_4+2 t_2 t_3 t_4+3 t_3^2 t_5+27 t_4^2 t_5+3 t_2 t_3 t_5+9 t_2 t_4 t_5+18 t_3 t_4 t_5
	  +4 t_3^2 t_6\notag\\
	  &+36 t_4^2 t_6+64 t_5^2 t_6+4 t_2 t_3 t_6 +12 t_2 t_4 t_6+24 t_3 t_4 t_6+16 t_2 t_5 t_6+32 t_3 t_5 t_6\notag\\
	  &+96 t_4 t_5 t_6.
\end{align}

\section{Discussion}

In this work, we investigated the relation of the threshold corrections for $\mc{N} = 2$, $D=4$ heterotic
string compactifications with Wilson lines
to Jacobi forms, where the Jacobi forms are over an even lattice and are possibly Jacobi forms of many variables. We showed 
that there are two kinds of Jacobi forms relevant in this context, a weight $-2$ Jacobi form
coming from $\Delta_\text{grav}$, 
$\phi_{-2,\lt}(\T, \vec{z})$, and a weight $0$ Jacobi form coming from $ 24 \Delta_\text{gauge} - \Delta_\text{grav}$,
$\psi_{0,\lt}(\T, \vec{z})$. The condition of being a Jacobi form is highly constraining since the vector space of Jacobi
forms over a lattice is finite dimensional. If one can determine some of the coefficients in Fourier expansions of 
$\phi_{-2,\lt}(\T, \vec{z})$ or $\psi_{0,\lt}(\T, \vec{z})$, even without any information about the full BPS spectrum it
may be possible to find out completely what these functions are using the finite dimensionality. We explored this idea
in a number of examples which are connected to an orbifold model in hypermultiplet moduli space and tested the results
using explicit computations of these Jacobi forms at the orbifold limit. An interesting future problem would be to generalize 
these methods to more general settings in which gauge symmetries of the low energy theory can arise.
One then can test how constraining the condition of
having Jacobi forms would be on the low energy effective theory. One should note that theories we consider are toroidal 
compactifications of $\mc{N}=1$, $D=6$ theories and finding constraints on such six dimensional
theories using their toroidal compactifications is similar in spirit to the work \cite{Seiberg:2011dr}.

We also computed threshold corrections and gave expressions for prepotential and gravitational coupling in terms of the 
Fourier coefficients of an appropriate Jacobi form. A detailed analysis of the Weyl chambers suggests extensions and
clarifications on \cite{Kawai:1998md}'s conjectures on mapping heterotic vector multiplet moduli to the vector multiplet moduli of a possible 
Type IIA dual. We studied this aspect using the examples we had on the heterotic side and worked out the resulting 
moduli mappings and cubic prepotentials. This side certainly deserves more attention to 
better understand the action of heterotic side's T-duality group in terms of more geometrical ideas on Calabi-Yau manifolds
and to test Gromov-Witten potentials obtained this way with explicit geometrical realizations.
A possible first step for this, which we worked out in detail on heterotic side,
may be the rank one case where the T-duality groups are extended paramodular groups, $\Gamma_m^+$, and associated Jacobi forms
are Jacobi forms in the sense of \cite{Zagier}.

\section*{Acknowledgement}
I am grateful to my advisor Jeff Harvey for his guidance and assistance along this project. I also acknowledge the support of 
NSF Grant 1214409.

\section*{Appendix: Definitions and Conventions}

In the main text and in the following we frequently use the notation $q$, $y$, $y_i$ where these mean
\begin{equation}
 q \equiv e^{2 \pi \ii \T} \text{, } y \equiv e^{2 \pi \ii z} \text{, and } y_i \equiv e^{2 \pi \ii z^i}.
\end{equation}

\subsubsection*{Basic Functions}
Our conventions for classical theta functions are as follows:
\begin{equation}
 \tht_1 (\T, z) = \ii \sum_{k \in \ZZ} (-1)^k q^{(k+1/2)^2/2} y^{k+1/2},
\end{equation}
\begin{equation}
 \tht_2 (\T, z) =  \sum_{k \in \ZZ}  q^{(k+1/2)^2/2} y^{k+1/2},
\end{equation}
\begin{equation}
 \tht_3 (\T, z) =  \sum_{k \in \ZZ}  q^{k^2/2} y^{k},
\end{equation}
\begin{equation}
 \tht_4 (\T, z) =  \sum_{k \in \ZZ} (-1)^k q^{k^2/2} y^{k}.
\end{equation}

The Dedekind eta function is defined as
\begin{equation}
 \eta (\T) = q^{1/24} \prod_{n=1}^{\infty} (1 - q^n).
\end{equation}

\subsubsection*{Polylogarithm Function}
Polylogarithm is defined by the infinite series
\begin{equation}
 \Li_s(z) = \sum_{n=1}^{\infty} \frac{z^n}{n^s},
\end{equation}
when $|z| < 1$. It can also be defined for $|z| \geq 1$ by analytic continuation.
We also use the function, $\mc{P}(z)$, as introduced in \cite{Harvey:1995fq}
\begin{equation}
 \mc{P}(z) = \Im\lp z \rp \Li_2 \lp e^{2 \pi \ii z} \rp + \frac{1}{2 \pi} \Li_3 \lp e^{2 \pi \ii z} \rp.
\end{equation}

\subsubsection*{Modular Forms}
A modular form of weight $k \in \ZZ$ is a holomorphic function
\begin{equation}
 \phi_k : {\mathbb H} \rightarrow \CC
\end{equation}
which satisfies the following two conditions where ${\mathbb H}$ is the complex upper plane:
\begin{itemize}
 \item For any $\left( \begin{array}{cc} a & b \\ c & d \end{array} \right) \in SL_2(\ZZ)$
      \begin{equation}
       \phi_k \lp \frac{a \T +b}{c \T + d} \rp =
			(c \T + d)^k \phi_k \lp \T \rp.
      \end{equation}
  \item $\phi_k$ has a Fourier expansion of the form
      \begin{equation}
       \phi_k \lp \T \rp = \sum_{n \in \ZZ} c(n) q^n,
      \end{equation}
      where $c(n)$ is zero unless $n \geq 0$. If, moreover, $c(0) = 0$ the modular form is called a cusp form.
\end{itemize}
The unique weight $12$ cusp form (up to an overall multiplicative constant) is
\begin{equation}
 \Delta (\T) = q \prod_{n=1}^{\infty} (1 - q^n)^{24} = q - 24 q^2 + 252 q^3 - 1472 q^4 + 4850  q^5 + \ldots.
\end{equation}
The ring of modular forms is freely generated by Eisenstein series $E_4$ and $E_6$ which are given by
\begin{equation}
 E_4 (\T) = 1 + 240 \sum_{n=1}^{\infty} \frac{n^3 q^n}{1 - q^n} =  1+ 240 q + 2160 q^2 + 6720 q^3 + 17520 q^4 + \ldots,
\end{equation}
and
\begin{equation}
 E_6 (\T) = 1 - 504 \sum_{n=1}^{\infty} \frac{n^5 q^n}{1 - q^n} =  1- 504 q - 16632 q^2 -122976 q^3 -532728 q^4 + \ldots.
\end{equation}
The Eisenstein series $E_2$ defined by
\begin{equation}
 E_2(\T) = 1 - 24 \sum_{n=1}^{\infty} \frac{n q^n}{1 - q^n} = 1 - 24 q - 72 q^2 - 96 q^3 - 168 q^4 + \ldots
\end{equation}
is not a modular form itself, however the non-holomorphic combination $ E_2(\T) - 3 / (\pi \Im \T)$ 
transforms under $SL(2,\ZZ)$ as if it is a weight $2$ modular form.

\subsubsection*{Jacobi Forms}
The theory of Jacobi forms is worked out in detail in \cite{Zagier}. In this work, we are using
a generalization of \cite{Zagier}, following
\cite{gritsenko1991jacobi, krieg1996jacobi, gritsenko1994modular, Kawai:2000px, choie2000differential,
clery2011modular, Gritsenko:2012qn}.

Let $L$ be a lattice endowed with a positive definite, symmetric and non-degenerate bilinear form
$ (,) : L \times L \rightarrow \ZZ$ and let $L$ be an even lattice with respect to this bilinear form.
By linearly extending this bilinear form, we also define the dual lattice
$L^*$, which consists of all elements of $L \otimes {\mathbb Q}$ having integral product with all elements of $L$.

A holomorphic (respectively weak or nearly holomorphic) Jacobi form of weight $k \in \ZZ$ 
associated with the lattice $L$ is a holomorphic function
\begin{equation}
 \phi_k : {\mathbb H} \times (L \otimes \CC) \rightarrow \CC
\end{equation}
which satisfies the following conditions:
\begin{itemize}
 \item For any $\left( \begin{array}{cc} a & b \\ c & d \end{array} \right) \in SL_2(\ZZ)$
      \begin{equation}
       \phi_k \lp \frac{a \T +b}{c \T + d}, \frac{\mathbf{z}}{c \T + d}\rp =
			(c \T + d)^k \exp{\lp  \frac{2 \pi \ii c (\mathbf{z},\mathbf{z})}{2(c \T + d)}  \rp}
			\phi_k \lp \T, \mathbf{z} \rp.
      \end{equation}
 \item For any $\mathbf{\lambda}, \mathbf{\mu} \in L $
      \begin{equation}
       \phi_k \lp \T, \mathbf{z} + \mathbf{\lambda} \T + \mathbf{\mu}\rp =
			  \exp{\lb - 2 \pi \ii \lp \frac{(\mathbf{\lambda},\mathbf{\lambda})}{2} +
				 (\mathbf{\lambda},\mathbf{z}) \rp \rb} \phi_k \lp \T, \mathbf{z} \rp.
      \end{equation}
  \item $\phi_k$ has a Fourier expansion of the form
      \begin{equation}
       \phi_k \lp \T, \mathbf{z} \rp = \sum_{\substack{ n \in \ZZ, \\ \mathbf{\alpha} \in L^*  }}
	      c(n, \mathbf{\alpha}) q^n \exp{\lp 2 \pi \ii  (\mathbf{\alpha},\mathbf{z}) \rp},
      \end{equation}
      where $c(n, \mathbf{\alpha})$ is zero unless $2n - (\mathbf{\alpha},\mathbf{\alpha}) \geq 0$ 
      (respectively unless $n \geq 0$ or unless $ n \geq -N$ for a positive integer $N$).
\end{itemize}
From the second property, one can show that Fourier coefficients depend only on the discriminant
$\Delta = n - (\mathbf{\alpha},\mathbf{\alpha})/2 $ and on $\alpha \text{mod }(L^*/L)$.

Another important property of Jacobi forms is that the space of weight k
holomorphic Jacobi forms for any $k$ and over an even lattice, $\lt$, is finite dimensional. This allows one to write the
most general weight $k$ Jacobi form over a lattice, $\lt$, and determine the whole function using only a few of its
Fourier coefficients.

In the main text, we usually choose a particular basis for $\lt$ and $\lt^*$ and then write $\phi_k$ in terms of this basis.
More explicitly, any $\mathbf{\alpha} \in L^*$ can be expanded as $\mathbf{\alpha} = k_i \vec{\gamma^i}$ and $\vec{z}$ can be
expanded as
$\vec{z} = z^i \vec{\beta_i} $, using the basis vectors $\{ \vec{\beta_i} \}$ of $\lt$ and its dual basis 
$\{ \vec{\gamma^i} \}$. Then, one can write
\begin{equation}
 \phi_{k}(\T, \z) = \sum_{n, k_i} c(n,k_i) q^n y_1^{k_1} \ldots y_s^{k_s},
\end{equation}
where the sum is over integers $n$ and $k_i$. We will indicate that a function is a weight k Jacobi form over the lattice
$\lt$ using subscripts $(k,\lt)$.

For $L = \langle 2 m \rangle$, one can check that the definition above reduces to the Jacobi from definition
of \cite{Zagier} where $m$ is called the index of the Jacobi form. To denote a weight $k$, index $m$ Jacobi form we use 
subscripts $(k,m)$. Two important examples are Eisenstein series $E_{4,1}(\T, z)$ and $E_{6,1}(\T, z)$ which have Fourier
expansions \cite{Zagier}:
\begin{equation}
 E_{4,1}(\T, z) = 1 + (y^2 + 56 y + 126 + 56 y^{-1} + y^{-2}) q + \ldots,
\end{equation}
and
\begin{equation}
 E_{6,1}(\T, z) = 1 + (y^2 - 88 y - 330 - 88 y^{-1} + y^{-2}) q + \ldots.
\end{equation}
Eisenstein series are 
constructed by starting with the constant, $1$, and then summing over all terms that can be obtained by acting on
$1$ with the members of the Jacobi group fixing the cusp at infinity. Therefore, in this sense, Eisenstein series 
comprise the simplest examples of Jacobi forms.

Let us denote the space of weight $k$, index $m$ weak Jacobi forms by $J^{\text{weak}}_{k,m}$. Then, an important structure
theorem in \cite{Zagier} tells that the ring of even weight weak Jacobi forms is freely generated by two weak Jacobi forms
$\tilde{\phi}_{-2,1}$ and $\tilde{\phi}_{0,1}$ over the ring of modular forms, where
\begin{align}
 \tilde{\phi}_{-2,1} = &- \frac{\tht_1(\T,z)^2}{ \eta (\T)^6} \notag\\
	  = &\frac{(y-1)^2}{y}-\frac{2 (y-1)^4 q}{y^2} 
	  +\frac{(y-1)^4 \left(y^2-8 y+1\right) q^2}{y^3} \notag\\
	  &+\frac{8 (y-1)^4 \left(y^2-3y+1\right) q^3}{y^3} \notag\\
	  &-\frac{(y-1)^4 \left(2 y^4-31 y^3+72 y^2-31 y+2\right) q^4}{y^4} + \ldots
\end{align}
and
\begin{align}
 \tilde{\phi}_{0,1} = &4 \lp \frac{\tht_2(\T,z)^2}{\tht_2(\T,0)^2} +\frac{\tht_3(\T,z)^2}{\tht_3(\T,0)^2}
			      + \frac{\tht_4(\T,z)^2}{\tht_4(\T,0)^2}\rp \notag\\
	= &\left(y+10+\frac{1}{y}\right)+\frac{2 (y-1)^2 \left(5 y^2-22 y+5\right) q}{y^2} \notag\\
	    &+\frac{(y-1)^2 \left(y^4+110 y^3-294 y^2+110 y+1\right) q^2}{y^3}  +\ldots.
\end{align}

Similarly, an explicit set of generators can be given for the case $\lt = m \lt_R$ where $m$ is a positive integer and $\lt_R$
is the root lattice of a simple Lie algebra (except for $E_8$). More explicitly, \cite{wirthmuller1992root} and
\cite{bertola1999jacobi} give generators for Weyl invariant Jacobi forms over $ \lt_R(m)$\footnote{$ \lt_R(m)$
is the lattice $ \lt_R$ rescaled by $\sqrt{m}$.} over the ring of
modular forms. Weyl invariance requires the Jacobi form to be invariant under the action of the Weyl group on $\vec{z}$.
Since in our physical examples states form irreducible representations of the gauge Lie algebra, Weyl invariance
condition is naturally satisfied. In the main text we give examples using generators for $A_2$ and $A_3$. To define those 
generators, it will be useful to define 
\begin{equation}
 \alpha(\T, z) = \ii \frac{\tht_1(\T,z)}{ \eta (\T)^3}
\end{equation}
and
\begin{equation}
 \beta(\T, z) = - \frac{1}{2 \pi} \frac{\del}{\del z} \lp \frac{\tht_1(\T,z)}{ \eta (\T)^3} \rp.
\end{equation}

For $A_2$, when we pick the lattice basis as
\begin{equation}
 \vec{\beta_1} = (1,-1,0) \text{ and }  \vec{\beta_2} = (0,1,-1),
\end{equation}
the generators of Weyl invariant Jacobi forms read 
\begin{align}
 \tilde{\phi}_{-3,A_2} = &\alpha(\T,z^1) \alpha(\T,z^2-z^1) \alpha(\T,-z^2) \notag\\
    = &\left(\frac{y_1}{y_2}-y_1+y_2-\frac{1}{y_2}-\frac{y_2}{y_1}+\frac{1}{y_1}\right)
    +\Big(-\frac{y_1^2}{y_2^2}+y_1^2+\frac{8 y_1}{y_2}-8 y_1-y_2^2+8 y_2 \notag\\
    &-\frac{8}{y_2} 
    +\frac{1}{y_2^2}-\frac{8 y_2}{y_1}+\frac{8}{y_1}+\frac{y_2^2}{y_1^2}-\frac{1}{y_1^2} \Big),
   q+\ldots
\end{align}
\begin{align}
 \tilde{\phi}_{-2,A_2} = & \beta(\T,z^1) \alpha(\T,z^2-z^1) \alpha(\T,-z^2) 
			    +\alpha(\T,z^1) \beta(\T,z^2-z^1) \alpha(\T,-z^2) \notag\\
			    &+\alpha(\T,z^1) \alpha(\T,z^2-z^1) \beta(\T,-z^2) \notag\\
	= &\left(-\frac{y_1}{2 y_2}-\frac{y_1}{2}-\frac{y_2}{2}-\frac{1}{2 y_2}
	    +3-\frac{y_2}{2 y_1}-\frac{1}{2 y_1}\right) \notag\\
	&+\Big(\frac{3 y_1^2}{y_2}-\frac{y_1^2}{2 y_2^2}-\frac{y_1^2}{2}+3 y_2 y_1-\frac{7 y_1}{y_2}
	  +\frac{3 y_1}{y_2^2}-7 y_1-\frac{y_2^2}{2}-7 y_2-\frac{7}{y_2}-\frac{1}{2 y_2^2} \notag\\
	  &+27+\frac{3 y_2^2}{y_1}-\frac{7 y_2}{y_1}+\frac{3}{y_2 y_1}-\frac{7}{y_1}-\frac{y_2^2}{2 y_1^2}
	  +\frac{3 y_2}{y_1^2}-\frac{1}{2 y_1^2}\Big) q + \ldots,
\end{align}
and
\begin{align}
 \tilde{\phi}_{0,A_2} = & 24 \mc{L}_{-2} \tilde{\phi}_{-2,A_2} \notag\\
      = &\left(\frac{y_1}{y_2}+y_1+y_2+\frac{1}{y_2}+18+\frac{y_2}{y_1}+\frac{1}{y_1}\right)
      +\Big(\frac{18y_1^2}{y_2}+\frac{y_1^2}{y_2^2}+y_1^2+18 y_2 y_1 \notag\\
      &-\frac{82 y_1}{y_2}+\frac{18 y_1}{y_2^2}-82 y_1 
      +y_2^2-82 y_2-\frac{82}{y_2}
      +\frac{1}{y_2^2}+378
      +\frac{18 y_2^2}{y_1}-\frac{82 y_2}{y_1} \notag\\
      &+\frac{18}{y_2 y_1}-\frac{82}{y_1}+\frac{y_2^2}{y_1^2}
      +\frac{18 y_2}{y_1^2}+\frac{1}{y_1^2}\Big) q+\ldots.
\end{align}
Note that the differential operator $\mc{L}_k$ is defined in equation \eqref{eq:JacobiDiff}.

For $A_3$, when we pick the lattice basis as
\begin{equation}
 \vec{\beta_1} = (1,-1,0,0) \text{, }  \vec{\beta_2} = (0,1,-1,0) \text{ and } \vec{\beta_3} = (0,0,1,-1),
\end{equation}
the generators of Weyl invariant Jacobi forms read 
\begin{equation}
 \tilde{\phi}_{-4,A_3} = \alpha(\T,z^1) \alpha(\T,z^2-z^1) \alpha(\T,z^3-z^2) \alpha(\T,-z^3),
\end{equation}
\begin{align}
 \tilde{\phi}_{-3,A_3} = &\beta(\T,z^1) \alpha(\T,z^2-z^1) \alpha(\T,z^3-z^2) \alpha(\T,-z^3) \notag\\
			    &+ \alpha(\T,z^1) \beta(\T,z^2-z^1) \alpha(\T,z^3-z^2) \alpha(\T,-z^3) \notag\\
			    &+\alpha(\T,z^1) \alpha(\T,z^2-z^1) \beta(\T,z^3-z^2) \alpha(\T,-z^3) \notag\\
			    &+\alpha(\T,z^1) \alpha(\T,z^2-z^1) \alpha(\T,z^3-z^2) \beta(\T,-z^3),
\end{align}
\begin{equation}
 \tilde{\phi}_{-2,A_3} = 24 \mc{L}_{-4} \tilde{\phi}_{-4,A_3},
\end{equation}
and
\begin{equation}
 \tilde{\phi}_{0,A_3} = 24 \mc{L}_{-2} \tilde{\phi}_{-2,A_3}.
\end{equation}

\bibliographystyle{JHEP.bst}
\bibliography{N2Jacobi}

\end{document}